\documentclass[usegraphicx,useAMS,usenatbib]{mn2e}
\usepackage{eso-pic}
\usepackage{placeins}
\usepackage[T1]{fontenc}
\usepackage{verbatim}
\usepackage[normalem]{ulem}
\usepackage{amsfonts,amsmath,amssymb}
\usepackage{times}
\usepackage[italic]{mathastext}
\usepackage{enumitem}
\usepackage[colorlinks,linkcolor=blue,citecolor=blue,urlcolor=blue,breaklinks]{hyperref}
\usepackage{color}
\usepackage{upgreek}
\usepackage{flushend}
\usepackage{lineno}
\usepackage{hyperref}
\usepackage{siunitx}
\usepackage{mathrsfs}

\usepackage[normalem]{ulem}

\newcommand{\redmapper}{redMaPPer}

\usepackage{todonotes}

\def\aap{A\&A}%
\def\aaps{A\&AS}%
\def\araa{ARA\&A}%
\def\mnras{MNRAS}%
\def\apj{ApJ}%
\def\apjl{ApJ}%
\def\apjs{ApJS}%
\def\aj{AJ}%
\def\pasj{PASJ}%
\def\physrep{Phys.~Rep.}%
\def\procspie{Proc.~SPIE}%

\AddToShipoutPictureBG*{%
  \AtPageUpperLeft{%
    \hspace{4\paperwidth}%
    \raisebox{-4.8\baselineskip}{%
      \makebox[0pt][l]{\textnormal{DES-2020-0629}}
}}}%

\AddToShipoutPictureBG*{%
  \AtPageUpperLeft{%
    \hspace{4\paperwidth}%
    \raisebox{-5.8\baselineskip}{%
      \makebox[0pt][l]{\textnormal{FERMILAB-PUB-21-049-AE}}
}}}%

\newcommand{\vect}[1]{{\boldsymbol#1}}

\title{Synthetic Galaxy Clusters and Observations Based on Dark Energy Survey Year 3 Data}

\author[T. N. Varga]{
\parbox{\textwidth}{
\Large
T.~N.~Varga,$^{1,2}$\thanks{corresponding author: \href{mailto:t.varga@campus.lmu.de}{t.varga@campus.lmu.de}}
D.~Gruen,$^{3,4,5}$
S.~Seitz,$^{1,2}$
N.~MacCrann,$^{6}$
E.~Sheldon,$^{7}$
W.~G.~Hartley,$^{8}$
A.~Amon,$^{4}$
A.~Choi,$^{9}$
A.~Palmese,$^{10,11}$
Y.~Zhang,$^{10}$
M.~R.~Becker,$^{12}$
J.~McCullough,$^{4}$
E.~Rozo,$^{13}$
E.~S.~Rykoff,$^{4,5}$
C.~To,$^{3,4,5}$
S.~Grandis,$^{14}$
G.~M.~Bernstein,$^{15}$
S.~Dodelson,$^{16}$
K.~Eckert,$^{15}$
S.~Everett,$^{17}$
R.~A.~Gruendl,$^{18,19}$
I.~Harrison,$^{20,21}$
K.~Herner,$^{10}$
R.~P.~Rollins,$^{21}$
I.~Sevilla-Noarbe,$^{22}$
M.~A.~Troxel,$^{23}$
B.~Yanny,$^{10}$
J.~Zuntz,$^{24}$
H.~T.~Diehl,$^{10}$
M.~Jarvis,$^{15}$
M.~Aguena,$^{25,26}$
S.~Allam,$^{10}$
J.~Annis,$^{10}$
E.~Bertin,$^{27,28}$
S.~Bhargava,$^{29}$
D.~Brooks,$^{30}$
A.~Carnero~Rosell,$^{31,26,32}$
M.~Carrasco~Kind,$^{18,19}$
J.~Carretero,$^{33}$
M.~Costanzi,$^{34,35,36}$
L.~N.~da Costa,$^{26,37}$
M.~E.~S.~Pereira,$^{38}$
J.~De~Vicente,$^{22}$
S.~Desai,$^{39}$
J.~P.~Dietrich,$^{14}$
I.~Ferrero,$^{40}$
B.~Flaugher,$^{10}$
J.~Garc\'ia-Bellido,$^{41}$
E.~Gaztanaga,$^{42,43}$
D.~W.~Gerdes,$^{44,38}$
J.~Gschwend,$^{26,37}$
G.~Gutierrez,$^{10}$
S.~R.~Hinton,$^{45}$
K.~Honscheid,$^{9,46}$
T.~Jeltema,$^{17}$
K.~Kuehn,$^{47,48}$
N.~Kuropatkin,$^{10}$
M.~A.~G.~Maia,$^{26,37}$
M.~March,$^{15}$
P.~Melchior,$^{49}$
F.~Menanteau,$^{18,19}$
R.~Miquel,$^{50,33}$
R.~Morgan,$^{51}$
J.~Myles,$^{3,4,5}$
F.~Paz-Chinch\'{o}n,$^{18,52}$
A.~A.~Plazas,$^{49}$
A.~K.~Romer,$^{29}$
E.~Sanchez,$^{22}$
V.~Scarpine,$^{10}$
M.~Schubnell,$^{38}$
S.~Serrano,$^{42,43}$
M.~Smith,$^{53}$
M.~Soares-Santos,$^{38}$
E.~Suchyta,$^{54}$
M.~E.~C.~Swanson,$^{18}$
G.~Tarle,$^{38}$
D.~Thomas,$^{55}$
and J.~Weller$^{1,2}$
\begin{center} (DES Collaboration) \end{center}
\bigskip
\small{{\em Author affiliations are listed at the end of this paper.}}
}}

\begin{document}
\date{\today}
\pagerange{\pageref{firstpage}--\pageref{lastpage}}
\pubyear{2021}
\maketitle

\label{firstpage}

\begin{abstract}
We develop a novel data-driven method for generating synthetic optical observations of galaxy clusters. In cluster weak lensing, the interplay between analysis choices and systematic effects related to source galaxy selection, shape measurement and photometric redshift estimation can be best characterized in end-to-end tests going from mock observations to recovered cluster masses. To create such test scenarios, we measure and model the photometric properties of galaxy clusters and their sky environments from the Dark Energy Survey Year 3 (DES Y3) data in two bins of cluster richness $\lambda\in[30;\,45)$, $\lambda\in[45;\,60)$ and three bins in cluster redshift ($z\in[0.3;\,0.35)$, $z\in[0.45;\,0.5)$ and $z\in[0.6;\,0.65)$. Using deep-field imaging data we extrapolate galaxy populations beyond the limiting magnitude of DES Y3 and calculate the properties of cluster member galaxies via statistical background subtraction. We construct mock galaxy clusters as random draws from a distribution function, and render mock clusters and line-of-sight catalogs into synthetic images in the same format as actual survey observations.  
Synthetic galaxy clusters are generated from real observational data, and thus are independent from the assumptions inherent to cosmological simulations. The recipe can be straightforwardly modified to incorporate extra information, and correct for survey incompleteness. New realizations of synthetic clusters can be created at minimal cost, which will allow future analyses to generate the large number of images needed to characterize systematic uncertainties in cluster mass measurements. 


\end{abstract}


\section{Introduction}
\label{sec:introduction}

The study of galaxy clusters has in recent years became a prominent pathway towards understanding the nonlinear growth of cosmic structure, and towards constraining the cosmological parameters of the universe \citep{Allen2011, KravtsovBorgani, Weinberg2012}. Weak gravitational lensing provides a practical method to study the mass properties of  clusters. It relies on estimating the gravitational shear imprinted onto the shapes of background source galaxies. The lensing effect is directly connected to the gravitational potential of the lens, and its measurement is readily scalable to an ensemble of targets in wide-field surveys \citep{Bartelmann01.1}. For this reason the lensing based mass calibration of galaxy clusters has become a standard practice for galaxy cluster based cosmological analyses \citep{rozoetal10,Mantz2015,planck_clusters_15, Costanzi19, Bocquet19, DESCosmo}.

Methods for estimating the shapes of galaxies include model fitting and measurements of second moments, with several innovative approaches developed in recent literature \citep{Zuntz13,Refregier13,Miller13,Bernstein14.1,HuffMETA,SheldonMETA, Metadetect}. 
Irrespective of the chosen family of algorithms, the performance of the shear estimates cannot be a-priori guaranteed, and needs to be validated in a series of tests (\citealp{Jarvis2016}, \citealp{FConti2017}, \citealp*{Y1shape}, \citealp{Samuroff18}, \citealp{Mandelbaum2018}, \citealp{KiDSshear}). These rely on synthetic observations: \emph{image simulations} which are then used to estimate the bias and uncertainty of the different methods in a controlled environment \citep{STEP2,GREAT08,GREAT3,Samuroff18, KiDSshear, Pujol, Y3sim}. 

Galaxy clusters present a unique challenge for validating weak lensing measurements for a multitude of reasons: they deviate  from the cosmic median line-of-sight in terms of the abundance and properties of cluster member galaxies \citep{Hansen08, ChunHao} resulting in increased blending among light sources (\citealp{Simet15blending, Euclid19,Eckert20}, \citealp*{Y3Balrog}), host a diffuse intra-cluster light (ICL) component \citep{Zhang19,Gruen2018_ICL, Santos2020, Kluge} influencing photometry, and induce characteristically stronger shear at small scales \citep*{rmy1}.

In this study we create synthetic galaxy clusters, and optical observations of these synthetic galaxy clusters in an unsupervised way from a combination of observational datasets. To achieve this, we measure and model the average galaxy content of \redmapper\ selected galaxy clusters in  Dark Energy Survey Year 3 (DES Y3) data along with the measurement and model for galaxies in the foreground and background. 
During this procedure the DES Y3 wide-field survey \citep{Y3gold} is augmented with information from deep-field imaging data \citep*{Y3deep}, resulting in enhanced synthetic catalog depth and better resolved galaxy features. 
Each synthetic cluster and its line-of-sight is generated as a random draw from a model distribution, which enables creating the large numbers of mock cluster realizations required for benchmarking precision measurements. This approach shortcuts the computational cost  and limited representation of reality of numerical simulations. 
The synthetic catalogs of cluster member galaxies and foreground and background galaxies along with the small-scale model for light around the cluster centers are then rendered into images in the same format as actual survey observations and can be further processed with the standard data reduction and analysis pipelines of the survey. 

The synthetic cluster images are controlled environments, where all light can be traced back to a source specified in the underlying model. A mass model calibrated by \citealp*{rmy1} is used to imprint a realistic lensing signal on background galaxies, which will enable future studies to perform end-to-end tests for recovering cluster masses from a weak lensing analysis of  synthetic images, incorporating photometric processing, shear and photometric redshift measurement and systematic calibration for lensing profiles and maps in a fully controlled environment. This is different from insertion based methods (\citealp{Suchyta2016}, \citealp*{Y3Balrog}), where synthetic galaxies are added onto real observations: Our method involves a generalization step avoiding re-using identical clusters multiple times, the full control of synthetic data allows quantifying the specific impact of the different cluster properties on the lensing measurement.

The primary focus of this work is to present the algorithm and a pilot implementation for generating synthetic cluster observations for the DES Y3 observational scenario mimicking the stacked lensing strategy of \citet*{rmy1} and \citet{DESCosmo}. Due to the transparent nature of the framework, changes and improvements aiming for increased realism: e.g. corrections for input photometry incompleteness or high resolution, deep cluster imaging, can be directly added to the model in future studies. For this reason, the presented algorithm is expected to be easily generalized and expanded to other ongoing (HSC: Hyper Suprime-Cam\footnote{http://hsc.mtk.nao.ac.jp/ssp/}, \citealp{HSC_Survey}; KiDS: Kilo-Degree Survey\footnote{http://kids.strw.leidenuniv.nl/index.php}, \citealp{KiDS2013}) and  upcoming (Vera C. Rubin Observatory\footnote{https://www.lsst.org/}, \citealp{LSST2019}; Euclid\footnote{http://sci.esa.int/euclid/}, \citealp{Euclid2011}; Nancy Grace Roman Space Telescope\footnote{https://wfirst.gsfc.nasa.gov/}, \citealp{WFIRST2015}) weak lensing surveys as well. 


The structure of this paper is the following: In \autoref{sec:desdata} we introduce the DES year 3 (Y3) dataset, in \autoref{sec:statmodel} we outline the statistical approach used in modeling the synthetic lines-of-sight, in \autoref{sec:model_results} we describe the concrete results of the galaxy distribution models derived from the DES Y3 dataset, and finally in \autoref{sec:random_los} we outline the method for generating mock observations for DES Y3.
In the following we assume a flat $\Lambda$CDM cosmology with $\Omega_{\rm m}=0.3$ and $H_0=70$ km s$^{-1}$ Mpc$^{-1}$, with distances defined in physical coordinates, rather than comoving.



\section{DES Y3 Data}
\label{sec:desdata}

\begin{figure}
    \centering
    \includegraphics[width=\linewidth]{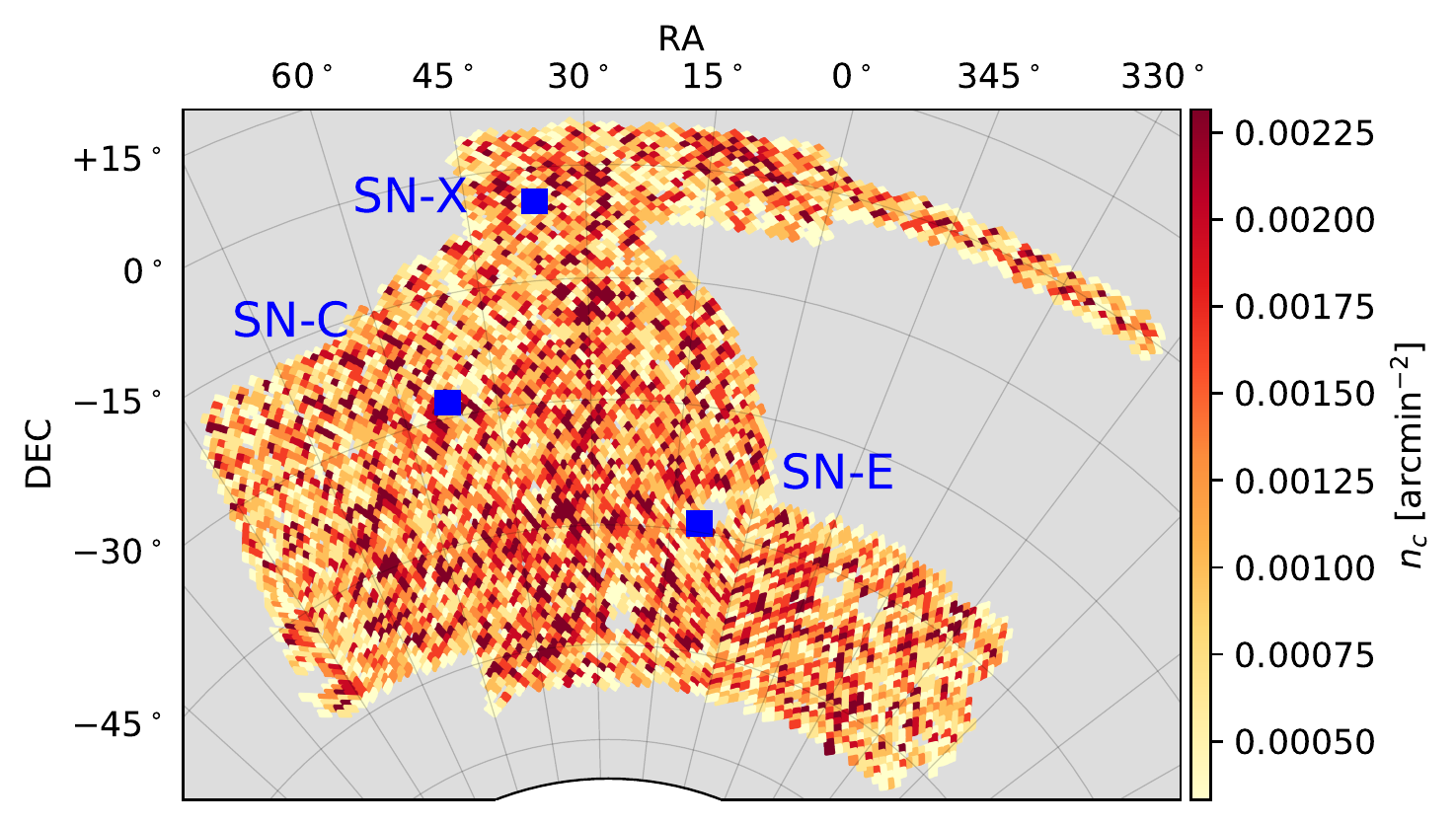}
    \caption{Footprint of targeted clusters in DES Y3. Blue markers: location of Deep field regions SN-C, SN-E, SN-X (marker size not to scale). The colorscale indicates the number density of galaxy clusters ($n_c$) identified by the \redmapper\ algorithm.}
    \label{fig:desy3_footprint}
\end{figure}
The first three years of DES observations were made between August 15, 2013 and February 12, 2016 \citep{DESmorethan, Y3gold}. This Y3 wide-field dataset has achieved nearly full footprint coverage albeit at shallower depth, with on average 4 tilings in each band ($g,r,i,z$) out of the eventually planned 10 tilings. From the full 5000 deg$^2$, the effective survey area is reduced to approximately 4400 deg$^2$ due to the masking of the Large Magellanic Cloud and bright stars. In parallel to the wide-field survey a smaller, deep field survey is also conducted covering a total unmasked area of 5.9 deg$^2$ in 4 patches \citep*{Y3deep}.  These consist of un-dithered pointings of the Dark Energy Camera (DECam, \citealp{Flaugher2015}) repeated on a weekly cadence, resulting in data 1.5 - 2 mag deeper than the wide-field survey. 
The DES Y3 footprint is shown on \autoref{fig:desy3_footprint}.
We use three of the four of DES Y3 Deep Fields denoted as SN-C, SN-E and SN-X, These consist of 8 partially overlapping tilings: three tilings for SN-C and SN-X, and two of the SN-E. Their location is also shown on \autoref{fig:desy3_footprint}. 

\subsection{Wide-field data}
\label{sec:des_wide}
The primary photometric catalog of DES Y3 is the Y3A2 GOLD dataset \citep{Y3gold}. This includes catalogs of photometric detections and parameters from the wide-field survey as well as the corresponding maps of the characteristics of the observations, foreground masks, and star-galaxy classification.

Data processing starts with single epoch images for which de-trending and photometric corrections are applied. They are subsequently  co-added to facilitate the detection of fainter objects. The base set of photometric detections is obtained via SExtractor \citep{Bertin96} from $r\,+\,i\,+\,z$ coadds. The fiducial photometric properties  for these detections are derived using the \emph{single-object-fitting} (SOF) algorithm based on the \texttt{ngmix} \citep{ngmix2015} software which performs a simultaneous fit of a bulge + disk composite model (CModel, \texttt{cm}) to all available exposures of a given object while modelling the point spread function (PSF) as a Gaussian mixture for each exposure. An expansion of this model is the \emph{multi-object-fitting} (MOF) \citep{Y3gold} approach where in addition to the above first step  friends-of-friends (FoF) groups of galaxies are identified based on their fiducial models, and in a subsequent step the galaxy models are corrected for all members of a FoF group in a combined fit. While for the Y3A2 GOLD dataset the SOF and MOF photometry were found to yield similar solutions, it is expected that in crowded environments the MOF photometry would perform better, due to its more advanced treatment of blending.

The 10$\sigma$ detection limit for galaxies using SOF photometry in the Y3A2 catalog is $g=23.78$, $r=23.56$, $i=23.04$, $z=22.39$  defined in the AB system \citep{Y3gold}. There is a  99 per cent completeness for galaxies with $i<22.5$. Star - galaxy separation is performed based on the morphology derived from SOF and MOF quantities, which for the $i<22.5$ sample has 98.5 per cent efficiency and 99 per cent purity,  yielding approximately 226 million extended objects out of a base sample of 390 million detections. SOF and MOF derived magnitudes are corrected for atmospheric and instrumental effects and for interstellar extinction to obtain the final corrected magnitudes.

\begin{figure}
    \centering
    \includegraphics[width=\linewidth]{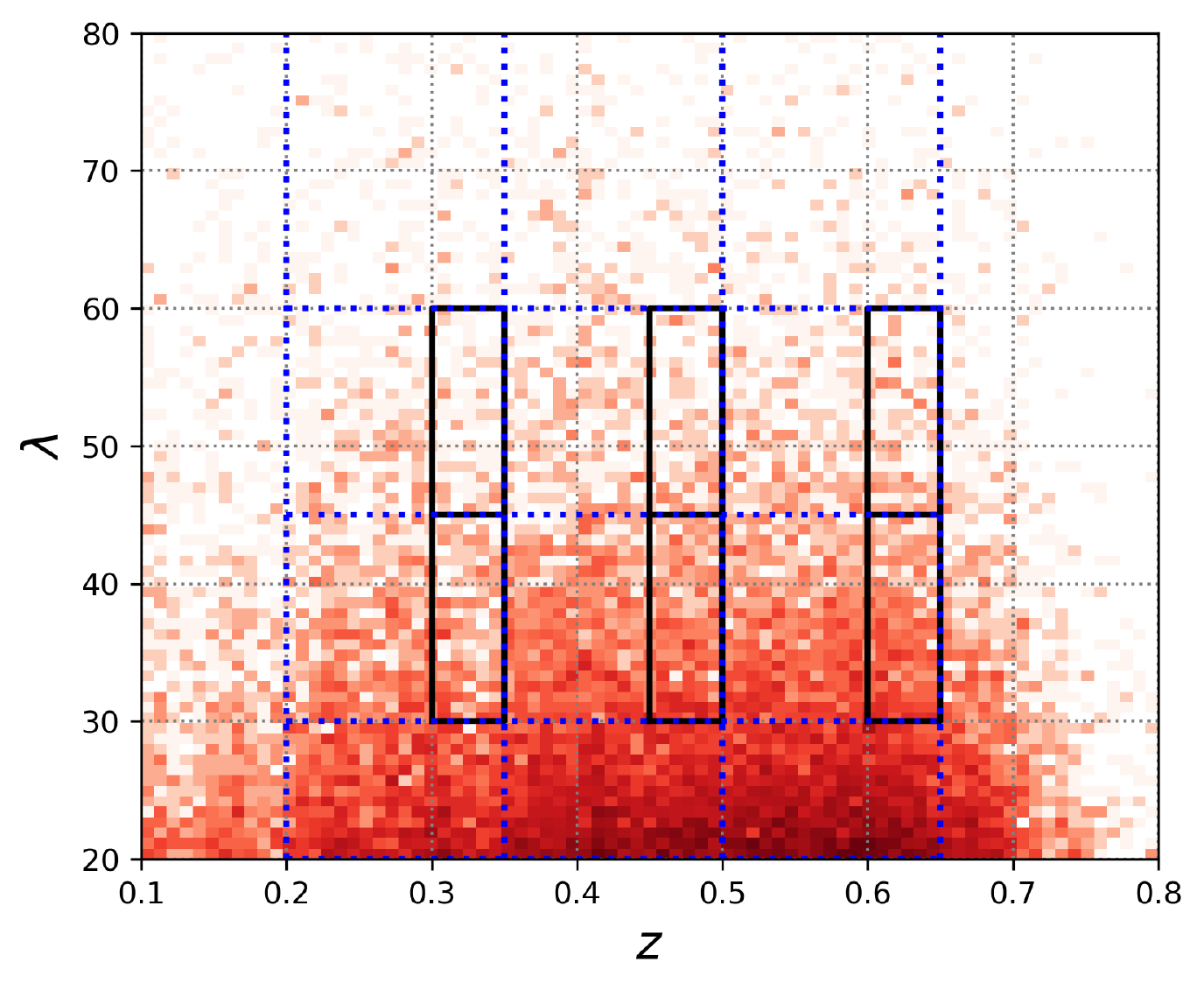}
    \caption[Distribution of richness and redshift values for redMaPPer clusters.]{Distribution of redMaPPer clusters in DES Y3 dataset in the volume-limited sample. \emph{Solid black rectangles:} narrow redshift selection. \emph{Blue dotted rectangles}: DES Y1 cluster cosmology selection.}
    \label{fig:rm_plot}
\end{figure}

 \begin{figure*}
    \small
    \centering
    \includegraphics[width=1.\linewidth]{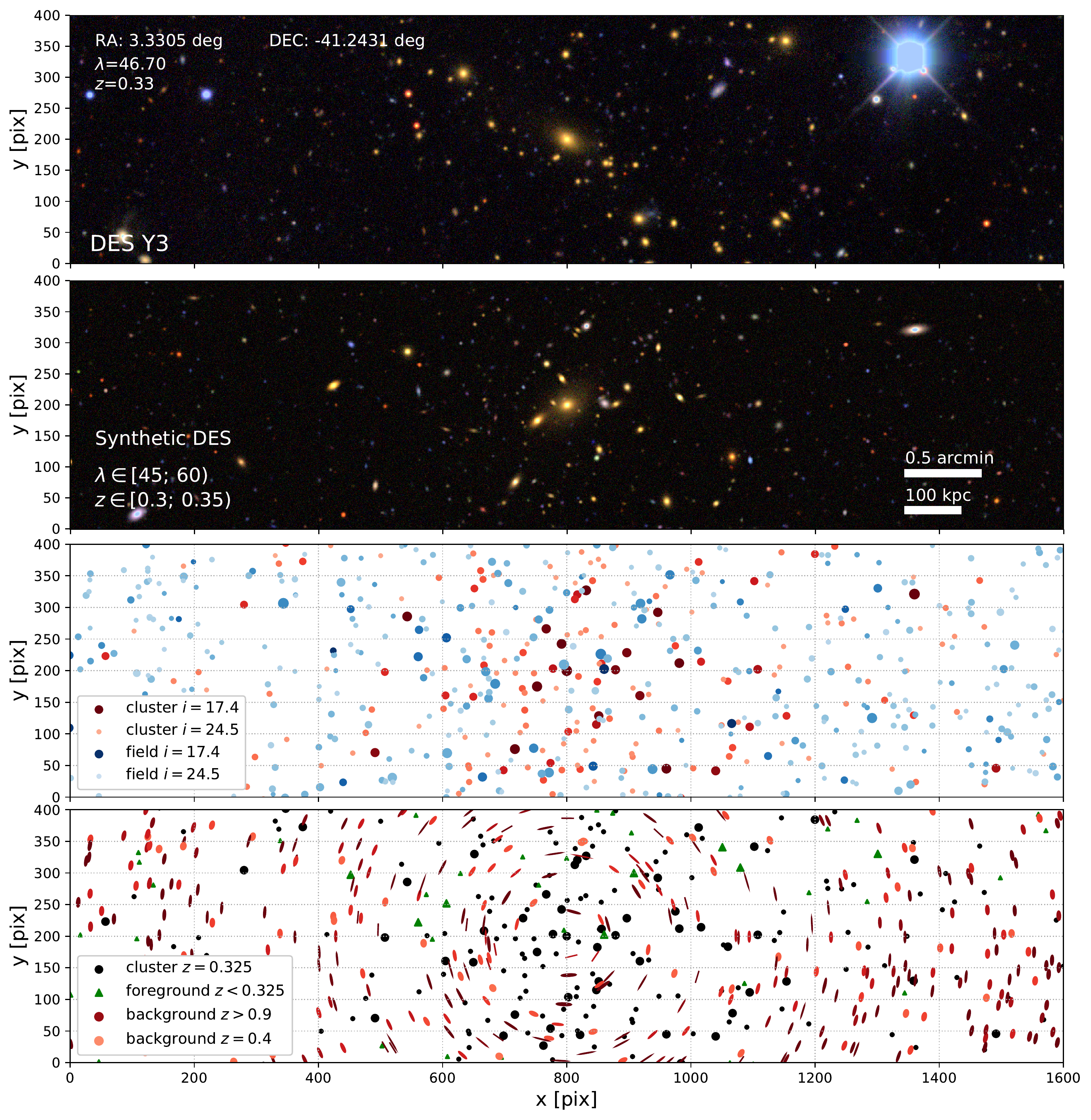}
    \caption[Synthetic galaxy cluster representative of $\lambda\in[45\;60)$, $z\in[0.3;\,0.35)$ and image of a real DES cluster in the same observable bin.]{Real and synthetic galaxy cluster side by side. \emph{Top:} $gri$ color composite image of a real \redmapper\ galaxy cluster in the DES Y3 footprint. \emph{Second row:} $gri$ color composite image of a synthetic galaxy cluster representative of $\lambda\in[45\;60)$, $z\in[0.3;\,0.35)$. \emph{Third row:} brightness distribution of the synthetic light sources for cluster members (red/brown) and foreground and background objects (blue). Darker shades and larger symbols correspond to brighter objects. \emph{Bottom row:} exaggerated shear map of background sources (red ellipses) with the shade representing redshift, cluster members (black) and foreground sources (green). }
    \label{fig:cluster_schematic}
\end{figure*} 

\subsection{RedMaPPer Cluster Catalog}
\label{sec:des_rm}

We consider an optically selected sample of galaxy clusters identified by the \redmapper\ algorithm in the DES Y3 data \citep{Rykoff2014_RM1}. The base input for this cluster finding is the Y3A2 SOF photometry catalog described above, from which \redmapper\ identifies galaxy clusters as overdensities of red-sequence galaxies.  This analysis uses \redmapper\ version v6.4.22+2. 
An optical mass proxy \emph{richness} $\lambda$ is assigned to each cluster defined by the effective number of red-sequence member galaxies brighter than $0.2\,L_*$. Cluster redshifts are estimated based on the photometric redshifts of likely cluster members yielding a  nearly unbiased estimate with a scatter of $\sigma_z / (1 + z) \approx 0.006$ (\citealp*{rmy1}).

We consider a locally volume-limited sample of clusters extending up to  $z\approx 0.65$, set by the survey completeness depth of $i\approx 22.6$. This \redmapper\ cluster catalog contains more than 869,000 clusters down to $\lambda>5$ and more than 21,000 above $\lambda>20$. The spatial distribution of the latter higher richness sample is shown on \autoref{fig:desy3_footprint}, and the richness and redshift distribution is shown on \autoref{fig:rm_plot}.  In addition to the cluster catalog, a catalog of reference random points is also provided, which are drawn from the part of the footprint where survey conditions permit the detection of a cluster of given richness and redshift. 

Finally we note that \redmapper\ uses SOF derived photometric catalogs instead of MOF, however this is expected to have no impact on the result of this work as we only utilize the positions, richnesses and redshifts of the clusters.

\subsection{Deep-Field Data}
\label{sec:des_deep}

The DES supernova and deep field survey is organized into four distinct fields: SN-S, SN-X, SN-C and SN-E (\citealp{Kessler2015, DESsn},\citealp*{Y3deep}). In this work we only consider the  SN-X, SN-C, SN-E fields covering a total unmasked area of 4.64 deg$^2$ which overlap with the VISTA Deep Extragalactic Observations (VIDEO) survey \citep{VIDEO}, providing $J,H,K$ band coverage.


In the present study we consider only the detections derived from the \textsc{COADD\_TRUTH} stacking strategy which aims to optimize for reaching approximately $10\times$ the wide-field survey depth while requiring that the deep field resolution (FWHM) be no worse that the median FWHM in the wide-field data \citep*{Y3deep}.

A difference compared to Y3A2 GOLD is that the MOF algorithm is run with "forced photometry" where astrometry and  deblending are done using DECam data, and  infrared bands incorporated only for the photometry measurement. This approach results in a coadded consistent photometric depth of $i=25$ mag. The photometric performance of these solutions were compared between the DES wide and deep field datasets using a joint set of photometric sources, finding very good agreement on the derived colors \citep*[see Fig. 12 of][]{Y3deep}. 
Additionally, for the deep field photometry the \texttt{ngmix} algorithm is run using the bulge + disk composite model with fixed size ratio between the bulge and disk components (in the following denoted as \texttt{bdf} to distinguish from the wide-field processing).

A photometric redshift estimate is derived by \citet*{Y3deep} for the deep-field galaxies via the \texttt{EAzY} algorithm \citep{eazy}. These photometric redshift estimates are obtained by fitting a mixture of stellar population templates to the $ugrizJHK$ band fluxes of the deep field galaxies. The possible galaxy redshifts and stellar template parameters are varied jointly to obtain a redshift probability density function. The redshift estimates are validated using a reference set of spectroscopic galaxy redshifts over the same footprint, and \citet*{Y3deep} finds overall good performance for bright and intermediate depths which however deteriorates into a very large outlier fraction for the faintest galaxies ($i$ > 24). In light of this we note that our algorithm for modeling the properties of cluster member galaxies presented in this analysis  does \textbf{not} rely on redshifts, and we consider photometric redshifts only for describing the line-of-sight distribution of foreground and background galaxies. Due to the substantially shallower limiting depth of the DES Y3 wide-field survey the impact of the increased fraction of very faint ($i$>24) redshift outliers is expected to be negligible.

\section{Statistical Model}
\label{sec:statmodel}

\subsection{Analysis Choices}

The focus of this study is to measure and model the galaxy content of \redmapper\ selected galaxy clusters within a bin of cluster properties, and to use this measurement to create mock galaxy clusters.  The cluster member model is complemented by a measurement and model for the properties of foreground and background galaxies. Each mock cluster is constructed to be representative in terms of its member galaxies of the whole bin of cluster properties, and does not aim to capture cluster-to-cluster or line-of-sight to line-of-sight variations.

By construction, the clusters identified by \redmapper\ are always centered on a bright central galaxy (BCG). Central galaxies form a unique and small subset of all galaxies, and therefore we treat them separately from non-central galaxies.
 In our synthetic observations we consider for each cluster bin a mock central galaxy which has the mean properties of the observed \redmapper\ BCG properties within that bin. In this study, we only consider clusters selected on richness and redshift (mimicking \citealp{DESCosmo}), and do not aim to incorporate correlated scatter between additional observables and mass properties at fixed selection. 
Thus the task for the rest of this section is  to model the properties and distribution of non-central, foreground and background galaxies, in the following simply denoted as galaxies. Faint stars are treated in the same framework as foreground galaxies, while bright stars, transients, streaks, and other imperfections which are masked during data processing are not incorporated in this model\footnote{Nevertheless, these can be added after the synthetic images are generated.}.

Throughout this analysis we assume that galaxies are to first order sufficiently described by a set of observable features, primarily provided by the DES photometric processing pipeline. The key features are: $i$-band magnitude $m_i$ with de-reddening and other relevant photometric corrections applied, colors $\vect{c}=(g-r,\,r-i,\,i-z)$,  galaxy redshift $z_g$, and morphology parameters $\vect{s}$ describing the scale radius, ellipticity and flux ratio of the two components of the \texttt{ngmix} SOF/MOF bulge + disk galaxy model. The full list of features and their relation to the DES Y3 data products is listed in \autoref{tab:feature_list}.

Our aim is to model the distribution of cluster member galaxies, and foreground and background galaxies in the space of the above features as a function of projected separation $R$ from galaxy clusters of richness $\lambda$ and redshift $z$.  These distributions cannot be directly measured from the DES wide-field survey as individual cluster member galaxies cannot be identified with sufficient completeness from photometric data alone, and the bulk of the galaxy populations lie beyond the completeness threshold magnitude of $i \approx 22.5$, where photometric errors come to dominate the derived features. To counteract this limitation we adopt a two-step approach: First a target distribution of well measured reference features, in this case a set of reference colors and radius $(\vect{c}_{ref};\, R\,|\, \lambda, z)$ is measured in the wide-field survey (\autoref{sec:indexer} and \autoref{sec:kde}). In the second step the wide-field target distribution is used as a prior for resampling the galaxy features measured in the DES Deep Fields (\autoref{sec:extrapolation}). Comparing the target distribution around clusters and around a set of reference random points enables us to isolate the feature distribution of cluster members (\autoref{sec:resampling}). Thus the resampling transforms the deep-field feature distribution into an estimate on the full feature distribution of cluster member galaxies, while keeping additional features measured accurately only in the deep-field data, and extrapolate the cluster population to fainter magnitudes. 

\autoref{fig:cluster_schematic} shows an illustration of a mock cluster generated as a result of this analysis at the level of a galaxy catalog and also as a fully rendered DES Y3-like coadd image, along with an actual \redmapper\ cluster taken from the DES Y3 footprint with similar richness and redshift. 

\subsection{Data Preparation}
\label{sec:indexer}

We group galaxy clusters into two bins of richness $\lambda\in[30;\,45)$ and $[45;\,60)$, and three bins of redshift $z\in[0.3;\,0.35)$, $[0.45;\, 0.5)$ and $0.6;\, 0.65)$, where each sample is processed separately. Our binning scheme is motivated by the selections of \citet*{rmy1} and  \citet{DESCosmo}, shown on \autoref{fig:rm_plot}. In this pathfinder study, however, we only cover their central richness bins, and enforce a narrower redshift selection to reduce the smearing of observed photometric features (e.g. red sequence) due to mixing of different redshift cluster members. While this smearing is not a limitation for the presented model, reduced smearing and redshift mixing will enable useful sanity checks in evaluating performance.

The base dataset for this study is a subset of the Y3A2 GOLD photometric catalog selected via the flags listed in \autoref{tab:y3mof_cuts}, queried from the DES Data Management system (DESDM, \citealp{DESDM}). The flags are chosen to yield a high-completeness galaxy sample while excluding photometry failures. For each cluster in a  given cluster selection we select all entries from this base catalog which are within a pre-defined search radius $\theta_\mathrm{query}\approx 6\, \mathrm{deg}$ around the cluster using the HEALPix algorithm \citep{Healpix}. 

Directly manipulating the above dataset is not feasible, therefore we select a weighted, representative subsample of entries: First we measure the total radial number profile of galaxies around the clusters in radial bins arranged as $[10^{-3};\, 0.1)$ arcmin, and in 50 consecutive logarithmically-spaced radial bins between 0.1 arcmin and 100 arcmin. Then, from each radial range we draw $N_{\rm draw} = \mathrm{min}(N_{\rm bin};\,N_{\rm th})$ galaxies where $N_{\rm bin}$ is the number of galaxies in the radial bin, and $N_{\rm th} = 10000$ is a threshold number. 

The random draws are equally partitioned across the $N_{\rm clust}$  clusters\footnote{That is from the vicinity of each cluster approximately $ N_{\rm draw} \, / \, N_{\rm clust}$ galaxies are drawn without replacement from each radial bin.}. To account for the number threshold $N_{\rm th}$, for each drawn galaxy a weight 
\begin{equation}
w_{\rm bin}=N_{\rm bin} / N_{\rm draw}
\label{eq:galweight}
\end{equation}
is assigned. Therefore the number of tracers representing the galaxy distribution is reduced in an adaptive way. For each selected galaxy the full catalog row is transferred from the GOLD catalog, and through the random draws the same galaxy can enter multiple times, but at different radii.

The outcome of the above is a galaxy photometry catalog  containing the projected radius $R$ of each entry measured from the targeted cluster sample with a weight for each entry. 
The measurement is repeated for a sample of reference random points selected in the same richness and redshift range as the cluster sample. This second dataset is representative of the \emph{field} galaxy distributions, however, through the spatial and redshift distribution of the reference random points it also incorporates the impact of survey inhomogeneities and masking.

Foreground stars appear in the projected vicinity of each galaxy cluster on the sky and also within the deep-field areas, and enter into the photometry dataset. The model presented in this study is not dependent on separation between stars and galaxies, as stars are automatically removed during statistical background subtraction. Nevertheless, the photometric properties of stars compared to galaxies increases the computational cost, as the difference between the proposal and target distribution increases when large number of stars are included. To counteract this we employ a size--luminosity cut $i-\mathrm{mag} < -50 + log_{10}(1 + T) + 22$ to remove the bulk of the stellar population\footnote{This simple size-luminosity cut was adopted as the DES deep field star galaxy separation was not yet finalized during the data preparation stage of this analysis. Any differences between that and the current form are expected to manifest only in the run time requirement of the rejection sampling step.}, where $T$ is the effective size of a detection defined as listed in \autoref{tab:feature_list}. These objects will be re-added at a later stage to produce survey-like observations.

\subsection{Kernel Density Representation of Survey Data}
\label{sec:kde}

Our aim is to generalize the features of a finite set of observed galaxies into an estimate on their multivariate feature probability density function (PDF). We  achieve this task via kernel density estimation (KDE), which is a type of unsupervised learning algorithm \citep{Parzen62, Hastie01}.  In brief, the finite set of data points are convolved with a \emph{Kernel function} $K(r, \,h)$, where $h$ is the bandwidth which sets the smoothing scale during the PDF reconstruction.
We adopt a multivariate Gaussian kernel function $K(r,\,h)$ formulated for $d$ dimensional data with a single bandwidth $h$ equal to the standard deviation.
This way gaps and undersampled regions are modeled to have non-zero probability. 
For the practical calculation of KDEs we make use of the \texttt{scikit-learn} implementation of the above algorithm\footnote{\url{https://scikit-learn.org/stable/modules/density.html}}. A benefit of this KDE implementation is that it is numerically optimized for large number of features, allowing for efficient future expansions, augmentations of the set of considered galaxy properties.

The photometry catalog  has features with very disparate scales\footnote{E.g., the value range and distribution of galaxy magnitudes and galaxy colors is markedly different.}. This means that any single bandwidth $h$ (smoothing scale) is not equally applicable for all dimensions. To address this we standardize and transform the input features before the KDE step into a set of new features which are better described by a single bandwidth parameter. First we subtract the mean of each feature, then perform a principle component analysis (PCA) to find the eigendirections of the input features \citep{Hastie01} via the \texttt{scikit-learn} implementation\footnote{\url{https://scikit-learn.org/stable/modules/decomposition.html}} and map the features of each galaxy into a set of eigenfeatures. Finally, these are standardized by dividing each eigenfeature by its estimated standard deviation among the sample. 

In order to find the optimal bandwidth $h$ for each KDE, we perform  k-fold leave-one-out cross-validation \citep{Hastie01}. Here the same base data is split into $k$ equal parts, and from these each part is once considered as the test data, and the remainder is used as the training data. In this approach the score $ \mathcal{S} = \frac{1}{N}\sum_j^N \mathrm{ln}\, p_n(x_j,\, h)\,$ is calculated $k=5$ times on  different training and test combinations, and from this a joint cross-validation score is estimated. The final KDE is then constructed from the full dataset using the bandwidth maximizing the cross-validation score.

Using PCA standardization, bandwidths can be expressed relative to the standard deviation $\sigma=1$ of the various standardized eigenfeatures. Based on this we evaluate the cross validation score on a logarithmically-spaced bandwidth grid from $0.01\sigma$ to $1.2\sigma$ for each KDE constructed. We find that $h=0.1\sigma$ simultaneously provides a good bandwidth estimate  for the deep-field and the wide-field KDEs, for this reason we adopt it as a global bandwidth for further calculations.

\subsection{Cluster and Field Population Estimates}
\label{sec:pop_estimates}

\begin{figure*}
    \centering
    \includegraphics[width=\linewidth]{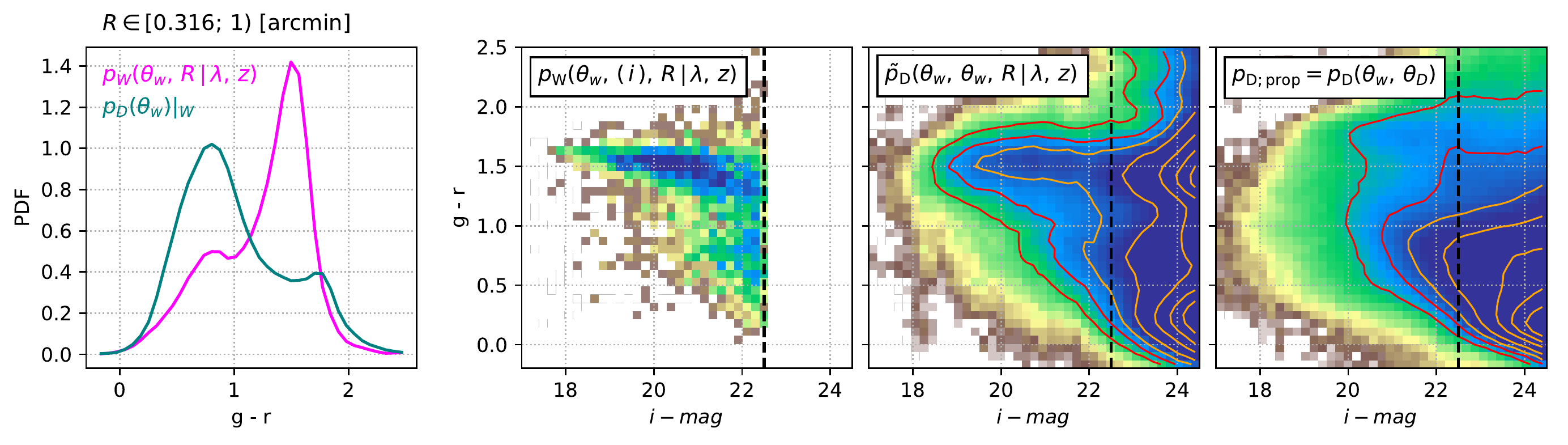}
    \caption{
Illustration of the re-weighting approach according to \autoref{eq:p_formula} and the various ingredients for the radial range $R\in[10^{-0.5};\, 1)$ arcmin around \redmapper~galaxy clusters with $\lambda\in[45;\,60$ and $z\in[0.3;\,0.35)$.  \emph{Left:} color PDF estimates for the wide-field shown in magenta, and the depth restricted Deep Field shown in green. \emph{Center left:} color-magnitude diagram of galaxies in the DES wide-field survey (not directly used in the transformation). This is the target which the transformation aims to reproduce for $i < 22.5$.
\emph{Center right:} transformed deep-field distribution according to \autoref{eq:p_formula}.
\emph{Right:} color-magnitude diagram of galaxies measured in the DES Deep Fields.
\emph{Dashed vertical lines:} wide-field completeness magnitude $i \approx 22.5$. The color scale and contour levels are identical in the three panels. For the $i<22.5$ magnitude range, the color based re-weighting shown on the center-right panel is in very good agreement with the color-magnitude distribution of the cluster line-of-sight shown on the center-left panel. The color scale is capped to the same level on the three right panels to allow direct comparison of the distributions.
    }
    \label{fig:res_illustration}
\end{figure*}

Our aim is to model the radial feature distribution of \emph{cluster member} galaxies for different samples of galaxy clusters. These must be separated from the distribution of foreground and background galaxies which we expect to be similar to the galaxies of the mean survey line-of-sight. The input data product for the following calculations is the feature PDF estimated from the various deep-field and wide-field galaxy catalogs for each using the KDE approach  in \autoref{sec:kde}. The full list of feature definitions are shown in \autoref{tab:feature_list}. 

Photometric redshift estimates available for the DES wide-field \citep*{Y1pz, Y3pz} are not precise  enough to isolate a sufficiently pure and complete sample of cluster member galaxies across the full range of galaxy populations (e.g. not only the red sequence). 
Therefore, to avoid the above limitation,  we perform a \emph{statistical background subtraction} \citep{Hansen08} to estimate the feature distribution of pure cluster member galaxies. In this framework we describe the line-of-sight galaxy distribution around galaxy clusters $p_{\rm clust}$ as a two-component system of a cluster member population $p_{\rm memb}$, and a field population which is approximated by the distribution around reference random points $p_{\rm rand}$. This yields
\begin{equation}
    p_{\rm memb}(\vect{\theta},\, R) = \frac{\hat{n}_{r}}{\hat{n}_{c} - \hat{n}_{r}} \left[\frac{\hat{n}_{c}}{\hat{n}_{r}} p_{\rm clust}(\vect{\theta},\, R) - p_{\rm rand}(\vect{\theta},\, R) \right]\,
    \label{eq:subtraction}
\end{equation}
where in practice both p.d.f-s on the right hand side are KDEs constructed from the wide-field dataset, $\vect{\theta}$ is the list of features considered, and $R$ is the projected separation from the targeted positions on the sky. $\hat{n}_{c}$ and $\hat{n}_{r}$ refer to the mean number of galaxies detected within $R_{max}$ around clusters and random points. 

The above approach is only applicable for those features $\vect{\theta}$ and their respective value ranges which are covered by the wide-field dataset. Furthermore, the formalism implicitly assumes that the p.d.f-s are dominated by the intrinsic distribution of properties, and not by measurement errors. To fulfill this requirement the wide-field data must be restricted to a parameter range where photometry errors play a subdominant role, and the completeness of the survey is high. 
This necessitates excluding the bulk of the galaxy population from the naive background subtraction scheme.

Especially important in relation to this study are galaxies whose flux is great enough to meaningfully contribute to the total light in a part of the sky, yet are not fully resolved or cannot be detected with confidence using standard survey photometry pipelines (\citealp{Suchyta2016}, \citealp*{Y3Balrog}). Nevertheless, these partial or non-detections have a significant impact on the photometric performance of survey data products \citep{Hoekstra17, Euclid19, Eckert20}. Therefore they must be modeled and included in the statistical description of a line-of-sight. A distinct undetected population of galaxies is associated with galaxy clusters, which are the faint-end of the cluster member galaxy population. The feature distribution of these galaxies is markedly different from the distribution of faint galaxies in the \emph{field} (cosmic mean) line-of-sight.

\subsection{Survey Depth and Feature Extrapolation}
\label{sec:extrapolation}

To characterize the properties of galaxies too faint to have complete detections in the DES wide-field survey, we make use of the DES Deep Fields.  Owing to significantly greater exposure time over many epochs, the completeness depth of the Deep Fields in the  \texttt{COADD\_TRUTH} mode is $\sim 2$ mag deeper than the Wide Fields \citep*{Y3deep}, and the measured fluxes and models of galaxy morphology are less impacted by noise at fixed magnitude compared to the DES Y3 GOLD wide-field catalog. Even for $i<22.5$ there are features measured more robustly for Deep Fields such as the \texttt{ngmix} SOF/MOF morphology model parameters. 
However, the colors of photometric sources detected in both datasets are found to be largely robust against the differences in the photometry analysis choices \citep*[see Section 2.3. of ][]{Y3Balrog}.  Therefore we aim to combine the galaxy distributions of the Deep Fields and the wide-field using colors to inform the extrapolation of the various feature distributions to fainter magnitudes. 

First, we denote our target distribution $p_D(\vect{\theta},\, R\,|\,\lambda,\, z)$, where the subscript $D$ indicates that the distribution is estimated from the Deep Fields down to a completeness limit of $i\approx24.5$. Similarly we denote distributions estimated from the wide-field dataset to the wide-field limiting magnitude with subscript $W$, and denote restricting a deep-field derived quantity to the shallower wide-field depth with $|_W$. \textit{In the following we decompose $\vect{\theta}$ into two sets of features: $\vect{\theta}_{\rm wide}$ which can be measured from the wide-field dataset, and $\vect{\theta}_{\rm deep}$ which can \textbf{only} be reliably measured from the Deep Fields:}
\begin{equation}
p_D(\vect{\theta},\,R\,|\,\lambda,\, z) \equiv p_D(\vect{\theta}_{\rm deep},\, \vect{\theta}_{\rm wide},\, R\,|\,\lambda,\, z)\,.
\label{eq:p_spit}
\end{equation}
Here we note that $R$, $\lambda$, and $z$ are features and quantities which also only originate from the wide-field dataset. We note that all features in $\theta_{\rm wide}$ can also be measured with confidence in the Deep Fields, but the reverse is not necessarily true.



Let us formulate \autoref{eq:p_spit} as a transformation of a naive proposal distribution:
\begin{align}
p_D(\vect{\theta}_{\rm deep},\, \vect{\theta}_{\rm wide},\, R\,|\,\lambda,\, z) =&\; p_{D:{prop\rm}}(\vect{\theta}_{\rm deep},\, \vect{\theta}_{\rm wide},\, R\,|\,\lambda,\, z)\\ \nonumber
& \times F(\vect{\theta}_{\rm deep},\, \vect{\theta}_{\rm wide},\, R\,|\,\lambda,\, z)\,.
\label{eq:factor_app}
\end{align}
Here we separate the task into two parts, where the proposal distribution $p_{D:\mathrm{prop}}$ carries information measured from the Deep Fields, and the multiplicative term $F$ represents the required transformation of the PDF. 
As there is no cluster information from the deep-field survey, the proposal PDF cannot depend on $\lambda$ and $z$:
\begin{align}
p_{D;\mathrm{prop}}(\vect{\theta}_{\rm deep},\, \vect{\theta}_{\rm wide},\, R\,|\,\lambda,\, z) = \,
& p_{D;\mathrm{prop}}(\vect{\theta}_{\rm deep},\, \vect{\theta}_{\rm wide},\, R)\,,
\end{align}
and for the same reason in the proposal distribution of $\vect{\theta}_{\rm deep}$ and $\vect{\theta}_{\rm wide}$ cannot be correlated with $R$:
\begin{align}
p_{D:\mathrm{prop}}(\vect{\theta}_{\rm deep},\, \vect{\theta}_{\rm wide},\, R\,|\,\lambda,\, z) = \,
& p_{D}(\vect{\theta}_{\rm deep},\, \vect{\theta}_{\rm wide}) \cdot p_{D:\mathrm{prop}}(R)\,.
\label{aeq:prop}
\end{align}
Here $p_{D}(\vect{\theta}_{\rm deep},\, \vect{\theta}_{\rm wide})$ can be directly measured from the deep-field survey, and $p_{D:\mathrm{prop}}(R)$ is chosen to capture the approximately uniform surface density of galaxies, e.g.  $p_{D:\mathrm{prop}}(R)\propto R$.

The remaining task is to find an appropriate multiplicative term  $F(\vect{\theta}_{\rm deep},\, \vect{\theta}_{\rm wide},\, R\,|\,\lambda,\, z)$ which transforms the proposal distribution $p_{D:\mathrm{prop}}$ into the target distribution $\tilde{p}_D$. \textit{In the following we denote with a tilde distributions or estimates which cover the full feature space, but are constrained by approximations due to information not accessible to us.} Since $\tilde{p}_D$ depends on $\lambda$, $z$ and $R$, and $p_{D:\mathrm{prop}}$ is independent of these, the $F$ term must contain all such information. Furthermore,  the correlation between $\vect{\theta}_{\rm deep}$ and $R$ cannot be measured from wide-field data, therefore we approximate $F$ as
\begin{equation}
\tilde{F}(\vect{\theta}_{\rm wide},\, R\,|\,\lambda,\, z)\approx F(\vect{\theta}_{\rm deep},\, \vect{\theta}_{\rm wide},\, R\,|\,\lambda,\, z) \,.
\end{equation}

A necessary consistency constraint placed on $\tilde{F}$ is  expressed as
\begin{align}
    \left.\tilde{p}_{D}(\vect{\theta}_{\rm wide},\, R\,|\lambda\,z)\right|_W &=\left.\, p_{D;\mathrm{prop}}(\vect{\theta}_{\rm wide},\, R)\right|_{W} \times \left. \tilde{F}(\vect{\theta}_{\rm wide},\, R\,|\,\lambda,\, z)\right.\label{eq:const1}\\
    &=\, p_{W}(\vect{\theta}_{\rm wide},\, R\,|\lambda,\,z)\,\label{aeq:const2}
\end{align}
where the $W$ subscript indicates a PDF estimated from wide-field data, and the $|_W$ subscript denotes that the otherwise greater magnitude range is restricted to the wide-field completeness magnitude of $i\approx22.5$. From the above constraint it is then possible to find the simplest form of $F$, as
\begin{align}
\tilde{F}(\vect{\theta}_{\rm wide},\, R\,|\,\lambda,\, z) &= \frac{1}{\hat{V}}\frac{p_{W}(\vect{\theta}_{\rm wide},\, R\,|\lambda,\,z)}{\left. p_{D;\mathrm{prop}}(\vect{\theta}_{\rm wide},\, R)\right|_W}\\
&= \frac{1}{\hat{V}}\frac{p_{W}(\vect{\theta}_{\rm wide},\, R\,|\lambda,\,z)}{\left. p_{D}(\vect{\theta}_{\rm wide}) \right|_W \cdot p_{D;\mathrm{prop}}(R)}\,
\label{aeq:F}
\end{align}
where $\hat{V}$ is a normalization factor to account for the different volumes of the wide-field and deep-field parameter spaces, e.g., the difference in the limiting depth of $i<22.5$ versus $i<24.5$.

From the combination of \autoref{aeq:prop} and \autoref{aeq:F} we can then write our estimate of the target distribution as
\begin{align}
\tilde{p}_D(\vect{\theta}_{\rm deep},\, \vect{\theta}_{\rm wide},\, R\,|\,\lambda,\, z) &\approx  \frac{p_{D}(\vect{\theta}_{\rm deep},\, \vect{\theta}_{\rm wide}) \,p_{W}(\vect{\theta}_{\rm wide},\, R\,|\lambda,\,z)}{\hat{V} \cdot \,\left. p_{D}(\vect{\theta}_{\rm wide}) \right|_W}\,,
\label{eq:p_formula}
\end{align}
where $p_{D;\mathrm{prop}}(R)$ drops out, and the approximation is composed entirely of p.d.f-s which can be directly measured from the wide-field or deep-field data.  
\textit{In simple terms, $p_{D}(\vect{\theta}_{\rm deep},\, \vect{\theta}_{\rm wide})$ describes the correlation between features seen only in the Deep Fields and features seen also in the wide-field survey, while $p_{W}(\vect{\theta}_{\rm wide},\, R\,|\lambda,\,z) \,/\, \left. p_{D}(\vect{\theta}_{\rm wide}) \right|_W$ captures the imprint of the cluster on the feature distributions}.
This framework conserves the color dependent luminosity function, and obeys
\begin{equation}
\tilde{p}_D(\vect{\theta}_{\rm deep}|\, \vect{\theta}_{\rm wide},\, R,\,\lambda,\, z)\equiv p_D(\vect{\theta}_{\rm deep}|\, \vect{\theta}_{\rm wide})\,.
\end{equation}\.
Since magnitudes are part of $\vect{\theta}_{\rm deep}$, this means that the final PDF estimate inherits the luminosity function of the Deep Fields, along with all additional features which are measured in the Deep Fields. 

An illustration of the outcome and the ingredients of this approach is shown on \autoref{fig:res_illustration}.  There, the center left panel shows the target distribution: the color-magnitude diagram of galaxies measured in projection with $R\in[10^{-0.5};\, 1)$ arcmin around \redmapper\ galaxy clusters with $\lambda\in[45;\,60$ and $z\in[0.3;\,0.35)$ in the DES wide-field survey. The leftmost panel shows a wide-field and the restricted deep-field feature (color) distribution. The rightmost panel shows the proposal distribution of galaxies measured in the DES Deep Fields, with the wide-field completeness magnitude shown as the vertical dashed line. The center right panel  shows the transformed deep-field distribution according to \autoref{eq:p_formula} where the radial color distribution around the cluster sample was used as the target PDF The color scale is identical in the three panels with  iso-probability contours overlayed.  For simplicity we take $\vect{\theta}_{\rm wide}= \vect{c}_{\rm wide}$ as a set of colors measured in both the wide-field survey and deep-field survey, and $\vect{\theta}_{\rm deep}= (m,\, \vect{s},\vect{c}_{\rm deep},\, z_g)$ is a vector composed of magnitudes, colors, morphology parameters and redshifts measured in the deep-field survey according to \autoref{tab:feature_list}.

\subsection{Rejection Sampling}
\label{sec:resampling}

\begin{figure*}
    \centering
    \includegraphics[width=\linewidth]{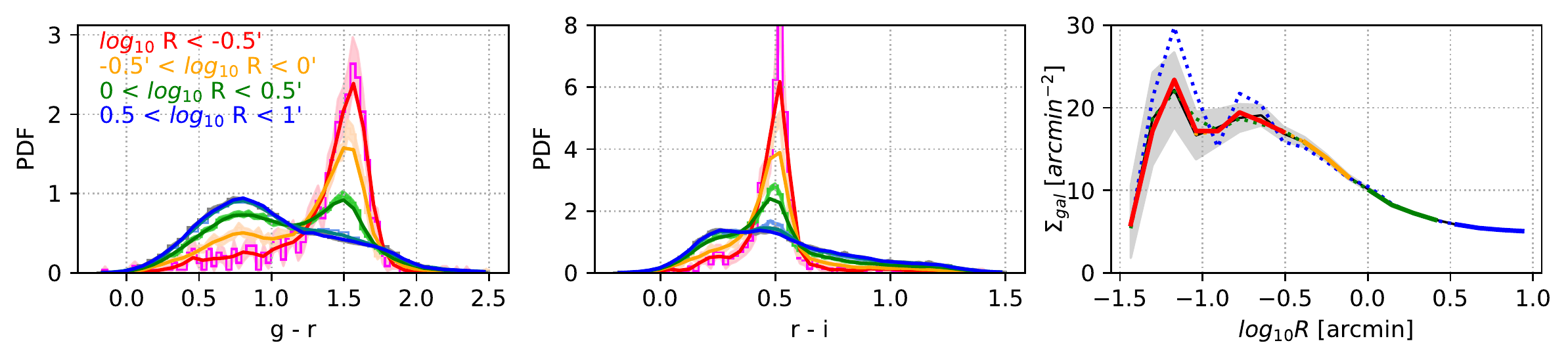}
    \includegraphics[width=\linewidth]{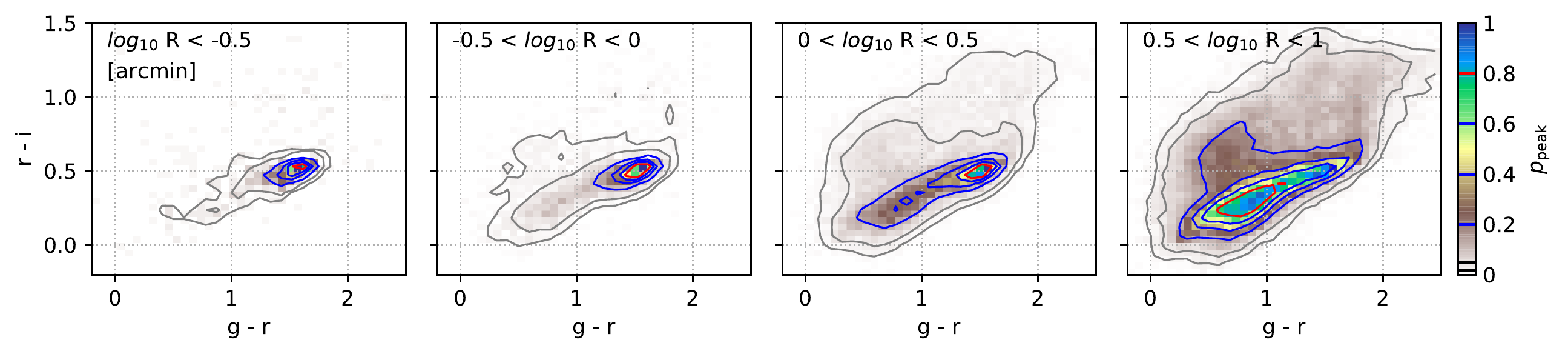}    
    \caption{Distribution of galaxy features with $i<22.5$ around \redmapper\ galaxy clusters ($\lambda\in[45;\,60)$, $z\in[0.3;\,0.35)$ in the DES wide-field dataset. 
    \emph{ Top left and center:} $g-r$ and $r-i$ color histograms of galaxies in bins of projected radius. \emph{Histogram:} DES data. \emph{Contours:} KDE reconstruction. The radial bins correspond to the radial shells used in the calculation.
    \emph{ Top right:} Surface density profile of galaxies around the targeted cluster sample. \emph{black:} measured profile. \emph{color:} KDE reconstruction of the surface density profile, color coded to the radial bins of the top left and center panels.
    \emph{Bottom:} $g-r$ - $r - i$ color distribution of galaxies in the four radial shells. Each panel is normalized to the same color and contour levels such that the broadening of the color distribution of galaxies and the reduction in the prominence of the red-sequence with increasing radius is clearly visible in the data and is well reproduced by the KDE.
    \emph{Histogram:} DES data. \emph{Contours:} KDE reconstruction. We note that the KDE is constructed globally for the full magnitude and feature ranges, and not only for the shown 2d marginal distribution.}
    \label{fig:input_cluster}
\end{figure*}

In the KDE framework, evaluating the PDF is computationally much more expensive than drawing random samples from it. Therefore, we adopt an approach where instead of directly performing the background subtraction we aim to generate random samples from the target distribution $\tilde{p}_{D;\, \mathrm{memb}}$.  For this we make use of an approach known as \emph{rejection sampling} \citep{MacKay02}. In short, this generates random variables distributed according to a \emph{target distribution} $p_{\rm targ}$ by performing random draws from a \emph{proposal distribution} $p_{\rm prop}$ which are then accepted or rejected according to a decision criterion.

\subsubsection{Background subtraction through resampling}
The cluster member galaxy population can be statistically defined as the feature dependent galaxy excess compared to a reference random line-of-sight shown in \autoref{eq:subtraction}. In the language of rejection sampling, $p_{\rm memb}$ can be calculated by stochastically estimating the volume between two PDFs \citep{MacKay02}. In our case the two distributions are $p_{rand}$ and $\frac{\hat{n}_c}{\hat{n}_r}p_{clust}$, the scaled feature PDF of galaxies measured in projection around reference random points and galaxy clusters respectively, and $\hat{n}_r$ and $\hat{n}_c$ refer to  the normalization factors, respectively. 

In the following we empirically sample $p_{\rm memb}$. For each sample:
\begin{enumerate}
    \item  Draw a proposal sample $\vect{\beta}_i\,\sim\,p_{\rm prop}\,\sim\,\mathcal{U}$, where $\beta_i$ is drawn from a uniform distribution whose support covers the support of both $p_{\rm clust}$ and $p_{\rm clust}$.
    
    \item Perform a uniform random draw $u_i\,\sim \mathcal{U}[0;\,1)$.
    
    \item Evaluate the acceptance condition
    \begin{equation}
   p_{\rm rand}(\vect{\beta}_i) < u_i \cdot\frac{\hat{n}_c}{\hat{n}_r} \, sup\left(p_{\rm clust}(\vect{\beta}_i)\right) < \frac{\hat{n}_c}{\hat{n}_r}p_{\rm clust}(\vect{\beta}_i) \,,
    \label{aeq:rs_subtr1}
\end{equation}
    and repeat from the previous step until the condition is fulfilled and a sample can be accepted. The rejection sampling recipe guarantees that  accepted samples will be distributed according to $p_{\rm memb}$. \citep{MacKay02}.
\end{enumerate}

Since in practice $p_\mathrm{clust}$ is not known exactly, we can rewrite Inequality \ref{aeq:rs_subtr1} by replacing it with an appropriately chosen value $\mathcal{M}$ which fulfills that $\frac{\hat{n}_c}{\hat{n}_r} p_\mathrm{clust} < \mathcal{M}$ and $p_\mathrm{rand} < \mathcal{M}$:
\begin{equation}
   p_{\rm rand}(\vect{\beta}_i) < u_i \cdot\frac{\hat{n}_c}{\hat{n}_r} \mathcal{M} < \frac{\hat{n}_c}{\hat{n}_r}p_{\rm clust}(\vect{\beta}_i)\,.
    \label{aeq:rs_subtr2}
\end{equation}
We further increase the acceptance rate by drawing samples $\vect{\beta}_i$ from an appropriately chosen proposal distribution $p_\mathrm{prop}$ instead of from a uniform distribution.
In this case the inequality modifies as
\begin{equation}
    \frac{p_{\rm rand}(\vect{\beta}_i)}{\frac{\hat{n}_c}{\hat{n}_r} \mathcal{M} \cdot p_\mathrm{prop}} < u_i < \frac{p_{\rm clust}(\vect{\beta}_i)}{\mathcal{M} \cdot p_\mathrm{prop}} \,
    \label{aeq:rs_subtr}
\end{equation}
where $\hat{n}_c / \hat{n}_r$ is the average relative overdensity of galaxy counts in the cluster line-of-sight compared to a reference random line-of-sight.

\subsubsection{Combining resampling and extrapolation}
\textit{The primary use of \autoref{aeq:rs_subtr} over directly performing the subtraction of the rescaled PDFs is that it can incorporate the extrapolation according to  \autoref{eq:p_formula}.} 
For this we adopt the proposal distribution as defined by \autoref{aeq:prop}:
\begin{align}
p_{\rm prop} &= p_{\rm prop}(\vect{\theta}_{{\rm deep}},\, \vect{\theta}_{\rm wide},\,R\,|\, \lambda,\, z)\nonumber \\
&= p_D(\vect{\theta}_{{\rm deep}},\, \vect{\theta}_{\rm wide}) \cdot p_{W;\,\mathrm{rand}}(R\,|\, \lambda,\, z)\nonumber\\
&= p_D(m,\, \vect{c},\, \vect{s},\, z_g) \cdot p_{W;\,\mathrm{rand}}(R\,|\, \lambda,\, z)\,,
\label{eq:p_prop_full}
\end{align}
which we use to draw the proposal random samples from. Furthermore, we define a restricted proposal distribution which contains only features contained within $\vect{\theta}_{ref}$, that is
\begin{align}
p_{rp} &= p_{rp}(\vect{\theta}_{\mathrm{wide}},\, R\,|\, \lambda,\, z)\nonumber\\
&= p_D(\vect{c}_{\mathrm{wide}}) \cdot p_{W;\,\mathrm{wide}}(R\,|\,\lambda,\, z)\,,
\label{aeq:p_prop_res}
\end{align}
which can be directly compared with $p_{clust}$ and $p_{rand}$.

Combining the above, we can generate random samples from the survey extrapolated $\tilde{p}_{memb}$, by drawing samples  $\{m_i,\, \vect{c}_i, \, \vect{s}_i,\, z_{g;i},\,R_i\}$ from \autoref{eq:p_prop_full}, and considering the subset which fulfills the extrapolated membership criteria
\begin{equation}
 \frac{\hat{n}_r}{\hat{n}_c}\frac{p_{W;\,\mathrm{rand}}(\vect{c}^\mathrm{ref}_{\rm wide;i},\,R_i\,|\,\lambda,\,z)}{\mathcal{M}\cdot p_{D}(\vect{c}^\mathrm{ref}_{\mathrm{wide};i})\cdot p_{W;\,\mathrm{rand}}(R_i\,|\,\lambda,\,z)} < u_i
\label{eq:rs_pmemb_resampling1}
\end{equation}
and
\begin{equation}
u_i< \frac{p_{W;\,\mathrm{clust}}(\vect{c}^\mathrm{ref}_{\mathrm{wide};i},\,R_i\,|\,\lambda,\,z)}{\mathcal{M}\cdot p_{D}(\vect{c}^\mathrm{ref}_{\mathrm{wide};i})\cdot p_{W;\,\mathrm{rand}}(R_i\,|\,\lambda,\,z)}\,.
\label{eq:rs_pmemb_resampling2}    
\end{equation}
Where $\vect{c}^\mathrm{ref}_\mathrm{wide}$ denotes a set of reference colors selected from $\vect{c}_\mathrm{wide}$: $\{g-r;\,r-i\}_{z1}$, $\{g-r;\,r-i\}_{z2}$ and $\{r-i;\,i-z\}_{z3}$ for the three cluster redshift bins respectively. These colors are chosen to bracket the red sequence at the respective redshift ranges in a manner similar to \cite{Rykoff2014_RM1}.

The above two inequalities define the decision criterion for the combined statistical background subtraction and extrapolation, and serve as the basis of the computation in this work.
Note that these criteria already implicitly contain the evaluation of \autoref{eq:p_formula} yielding an estimate of $\tilde{p}_{\rm memb}$, and are composed entirely of factors which can be directly estimated from either the wide-field or the deep-field galaxy datasets.

As a null-test, we can also perform the same resampling for the galaxies around random points, which using the same proposal distribution as above, is defined by the criterion
\begin{align}
u_i < \frac{\hat{n}_r}{\hat{n}_c}\frac{p_{W;\,\mathrm{rand}}(\vect{c}^\mathrm{ref}_{\mathrm{wide};i},\,R_i\,|\,\lambda,\,z)}{\mathcal{M}\cdot p_{D}(\vect{c}^\mathrm{ref}_{\mathrm{wide};i})\cdot p_{W;\,\mathrm{rand}}(R_i\,|\,\lambda,\,z)}\,,
\label{eq:rs_pfield_resampling}
\end{align}
which generates samples from the extrapolated field galaxy distribution $\tilde{p}_{rand}$.

In the above formulas the factor $\mathcal{M}$ must be chosen appropriately to ensure that the ratios are always less than or equal to unity. In practice there is no  recipe for $\mathcal{M}$, and the suitable value must be found for the actual samples proposed. Furthermore, measurement noise leads to small fluctuations in the KDEs which especially in the wings of the distributions manifests as $p_{\rm targ} / p_{\rm prop}$ being very poorly constrained. To regularize this behaviour we relax the requirement on $\mathcal{M}$ and in practice only require the criterion to be fulfilled for 99 per cent of the proposed points. We explore the $\mathcal{M}$ range in an iterative fashion up to 500, and find no significant change in the distribution of the samples for $\mathcal{M}>40$, thus  we adopt $\mathcal{M}=100$ throughout this study.

The random draws can be repeated until a sufficiently large sample is accepted for the cluster member and the field object dataset. Accepted draws can either be used directly to construct mock observations, or alternatively a KDE can then be constructed to estimate the PDF of the cluster members and extrapolated field galaxies separately.

A practical limitation of this sampling method is that since the proposal $R_i$ values are drawn from the full considered radial range around clusters and reference random points, the larger radial ranges will be much better sampled than the lower radius ranges because of the increase in surface area.
In our implementation we counteract this by simultaneously considering multiple nested shells of overlapping radial intervals to ensure the efficient covering of the full radial range. While each of these PDFs is individually normalized to unity, we express the relative probability $p_l$ of a member galaxy residing  in a given radial interval $r_l$ around a cluster as
\begin{equation}
    p_l \approx \left. \frac{\hat{n}_{c;\,l} - \hat{n}_{r;\,l}}{p_l(i <22.5)} \right/ \sum_l  \frac{\hat{n}_{c;\,l} - \hat{n}_{r;\,l}}{p_l(i <22.5)}\,
    \label{eq:nested}
\end{equation}
where $\hat{n}_{c;\,l}$, $\hat{n}_{r;\,l}$ is the average number of galaxies around clusters and random points residing in the radial bin in the wide-field dataset, and $p_l(i<22.5)$ is the probability that based on the  KDE in radial bin $l$ a galaxy is bright enough to be in the wide-field selection. While this formalism is similar to the direct background subtraction scheme defined in \autoref{sec:pop_estimates}, it is only used to approximate the relative weight of different radial ranges, and does not influence the estimation of the feature PDFs within the radial ranges.

\section{Model Results}
\label{sec:model_results}

\subsection{Input Feature KDEs}

For each sample of galaxy clusters we present the  measurements and the corresponding KDE estimates for the two primary input distributions: The distribution of features around clusters in the wide-field data, and the distribution of features in the deep-field dataset. 
\textit{We note that each KDE is constructed globally for all features and the full value range, and not only for the shown conditional distributions.}

\subsubsection{Distributions of wide-field galaxies around clusters}

\label{sec:result_clust}
\autoref{fig:input_cluster}  shows the measured feature distribution of galaxies around a selection of \redmapper\ galaxy clusters with $\lambda\in[45;\,60)$ and $z\in[0.3;\,0.35)$. The features of this distribution are the reference colors $\vect{c}_{ref} = (g-r,\,r-i)$ and the projected radial separation $R$ measured from the target galaxy cluster centers. Using these sets of features a KDE is constructed according to \autoref{sec:kde}, whose model for the PDF is shown as the continuous curves and contours on \autoref{fig:input_cluster}, while the 1D and 2D histograms represent the measured data. 

\begin{figure}
    \centering
    \includegraphics[width=\linewidth]{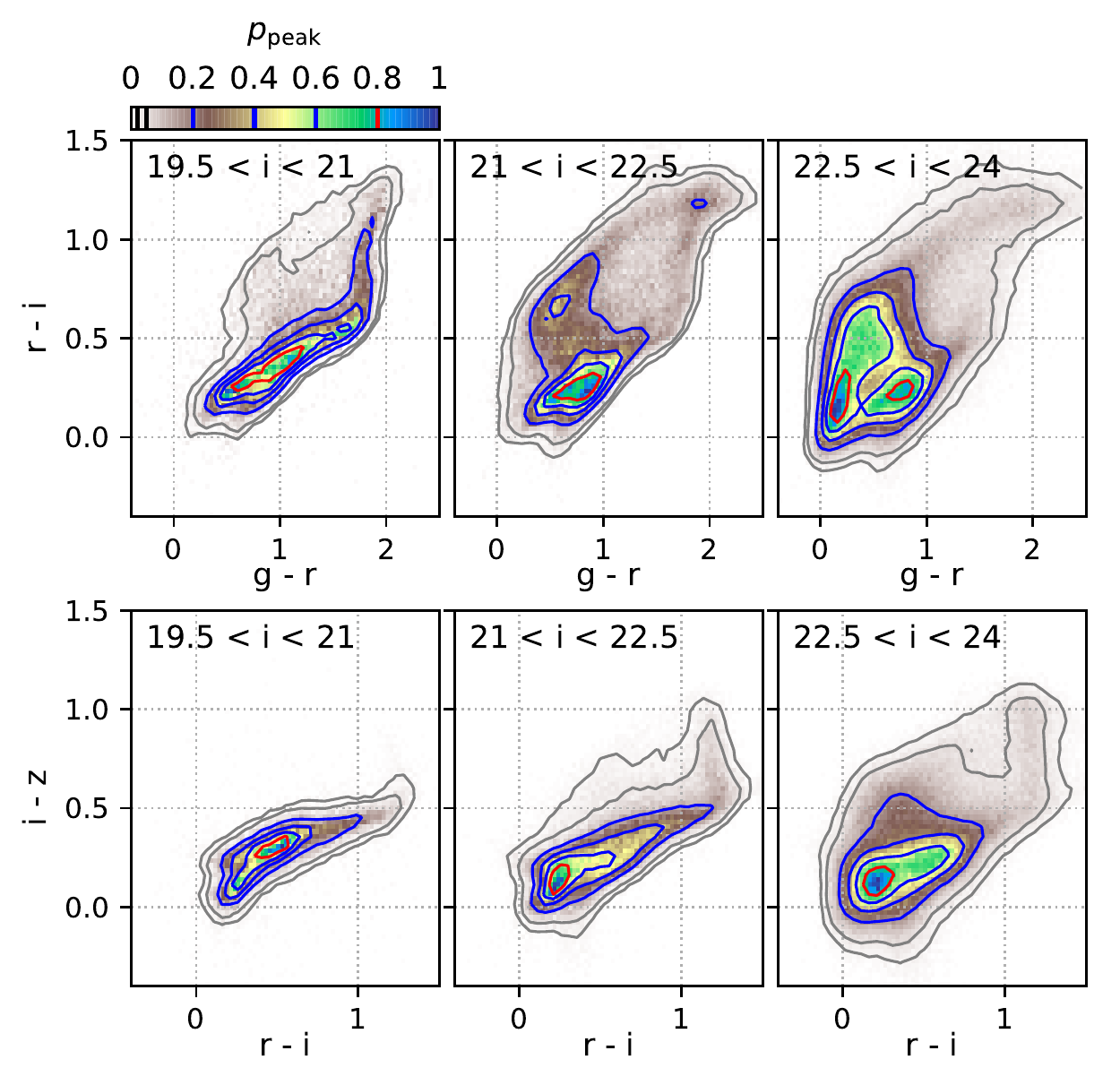}
    \caption{Distribution of $g-r$, $r-i$, $i-z$ galaxy colors in the DES Deep Fields in bins of $i$-band magnitude. \emph{Histogram:} DES data. \emph{Contours:} KDE reconstruction. We note that the KDE is constructed globally for the full magnitude and feature ranges, and not only for the shown 2d marginal distribution.}
    \label{fig:input_deep1}
\end{figure}

\begin{figure}
    \centering
  \includegraphics[width=\linewidth]{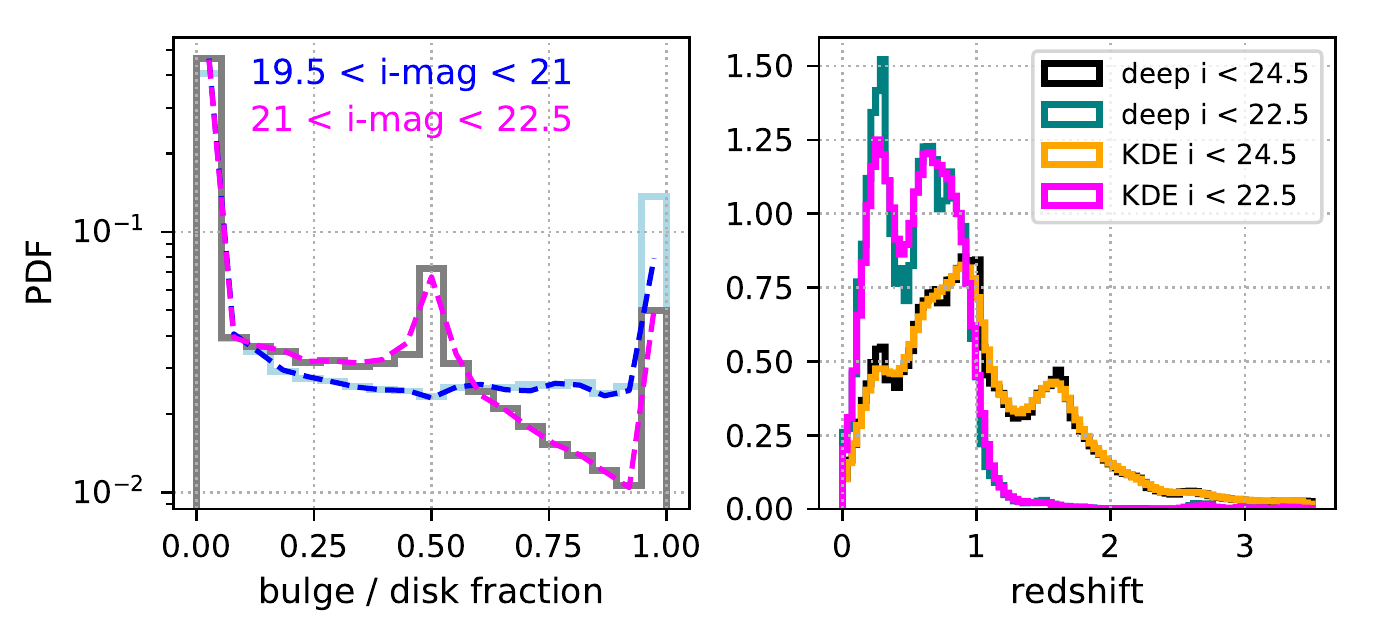}
    \caption{Distribution of galaxy morphology parameters in the DES Deep Fields, as listed in \autoref{tab:feature_list}.  \emph{Histogram:} DES data. \emph{Contours / curves:} KDE reconstruction. We note that the KDE is constructed globally for the full magnitude and feature ranges, and not only for the shown marginal distributions.
    }
    \label{fig:input_deep3}
\end{figure}

The top left two panels of \autoref{fig:input_cluster} show galaxy colors at different projected radii from the cluster center for all galaxies with $i< 22.5$, while the bottom panels  show the $g-r$ - $r-i$ color-color diagram of galaxies with $i<22.5$ in different radial bins. The histograms correspond to the measured distributions, while the contours represents the appropriate slice of the global KDE model. A prominent radial dependence is visible as the red sequence becomes increasingly dominant for small radii. The KDE model provides a good overall description of these galaxy distributions capturing the two-component nature of the galaxy population. It recovers the position and the approximate relative weight of the red sequence population.
We note that since the targeted galaxy clusters span a redshift range $\Delta z = 0.05$, the width of the observed red sequence population is measured to be wider, by this dispersion, compared to its intrinsic width.


The top right panel of \autoref{fig:input_cluster} shows the surface number density profile $\Sigma_{\rm gal}(R) = N(R) \,/\, 2\pi R$ of galaxies with $i<22.5$ around the selected cluster sample in the wide-field survey as the solid black curve. Colored curves show the corresponding KDE models for the four nested shells. In addition to the target range of the KDEs which are shown as the full lines, as a consistency test the interior continuation of the KDE model for the outermost nested spherical bin is shown as the dotted line. This only shows mild deviation from the respective profile of the data, and the measured radial \emph{surface density} profile and the KDE models show very good agreement. This means that the difference between the measured and modeled absolute density is very small over a range of two orders of magnitude, as set by the change in area element.

\begin{figure*}
    \centering
    \includegraphics[width=\linewidth]{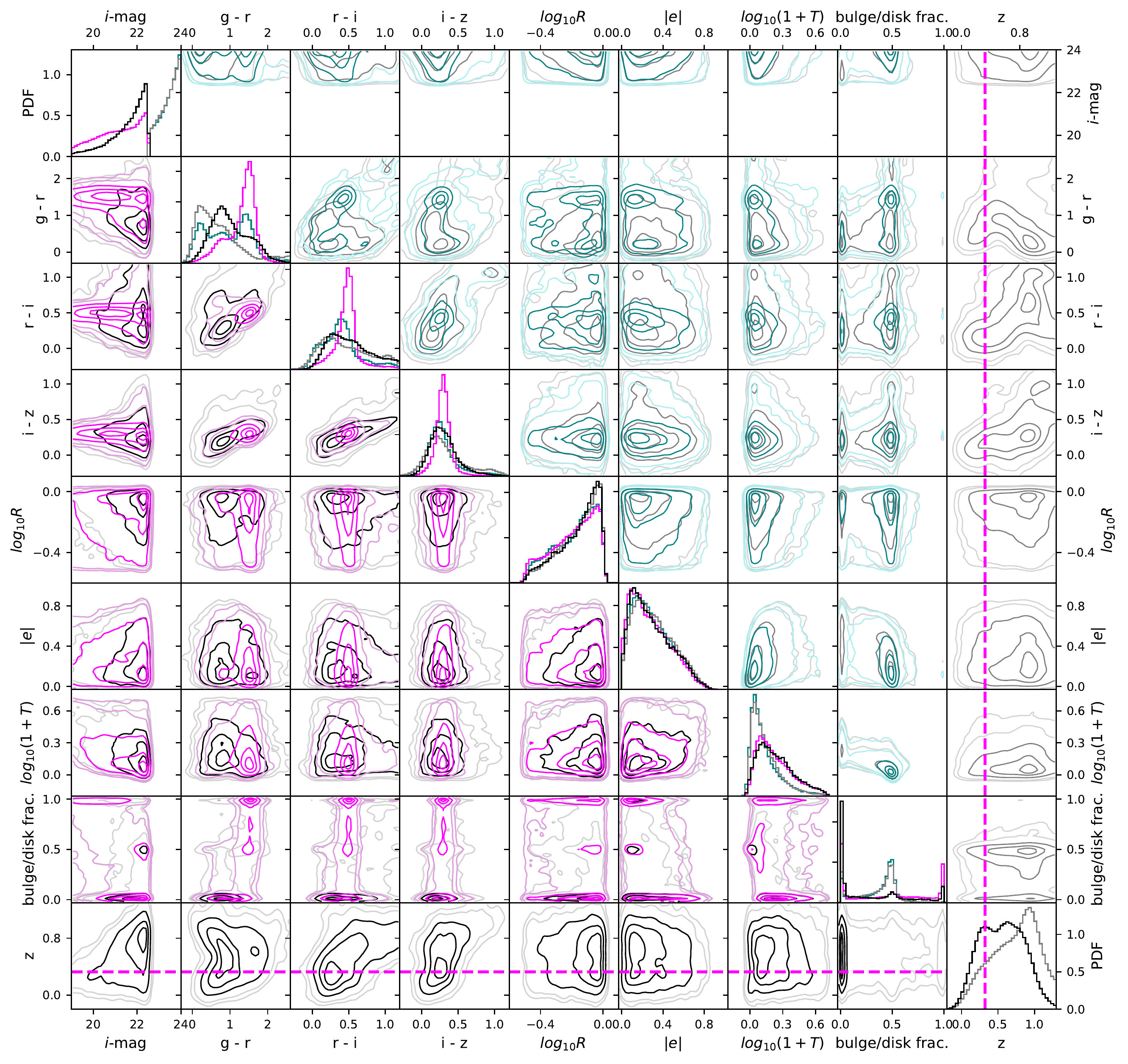}
    \caption{Joint galaxy feature model in the radial range $R\in[10^{-0.5};\,1]$ arcmin, for the cluster sample with  $\lambda\in[45;\,60)$ and $z\in[0.3;\,0.35)$. The parameters shown are summarized in \autoref{tab:feature_list}. \emph{Lower left panels, magenta:} cluster member galaxies with $i<22.5$. \emph{Lower left panels, black:} field galaxies with $i<22.5$. \emph{Upper right panels, green:} extrapolated cluster member galaxies  $22.5 < i < 24$. \emph{Upper right panels, gray:} extrapolated foreground and background galaxies with  $22.5 < i < 24$. The bump visible in the redshift PDF near the cluster redshift range (magenta dashed lines) is coincidental, it is a property of the DES deep-field galaxy distribution, also visible on \autoref{fig:input_deep3}.}
    \label{fig:joint model}
\end{figure*}

\subsubsection{Distributions of deep-field galaxies}

\label{sec:result_deep}

\autoref{fig:input_deep1} shows the $g-r$ - $r-i$ and the $r-i$ - $i-z$ color - color diagrams of the deep-field galaxies in three different magnitude ranges. The measured distributions are shown as a 2D histograms, and the corresponding KDE model is represented by contours. This KDE model is constructed simultaneously for all features listed in \autoref{tab:feature_list}, and it provides an excellent description of the color-color-magnitude distribution of galaxies. 

\autoref{fig:input_deep3} shows the same KDE model projected into the space of bulge / disk flux fraction (a morphology parameter) and redshift estimate. The left panel of \autoref{fig:input_deep3} shows the histograms of the measured bulge / disk flux fraction of the \texttt{ngmix} \texttt{bdf} galaxy model for two magnitude bins $19.5<i<21$ and $21<i<22.5$, along with the corresponding KDE model. Brighter galaxies are more likely to be bulge dominated (e.g. described by a de Vaucouleurs light profile) compared to fainter galaxies, which is in accordance with expectations from galaxy evolution \citep{Gavazzi}. The peak appearing at 0.5 is an imprint  of the morphology prior of the deep-field photometry pipeline, and it becomes prominent for the fainter galaxy selection as there the available information to constrain morphology from survey observations diminishes. KDE estimates cannot reproduce the hard cutoff edges $[0;\,1]$ of the bulge / disk flux fraction value, and for this reason we cap the distributions around 0 and 1 to restrict the PDF model to the appropriate interval, so that values greater than 1 or lower than 0 receive a value of 1 or 0 respectively.
The right panel of \autoref{fig:input_deep3} shows the estimated redshift distribution of the deep-field galaxies, as predicted by the EAzY algorithm (\citealt{eazy}, see \autoref{sec:des_deep}) along with the KDE reconstruction for two different magnitude ranges. For both the bulge/disk ratio and the redshift parameters the KDE model provides a very good description of the measured data. \textit{We emphasize that these are different projections of the same model shown on \autoref{fig:input_deep1}.}

\subsection{Cluster Member Feature Distributions}

\begin{figure}
    \centering
    \includegraphics[width=0.9\linewidth]{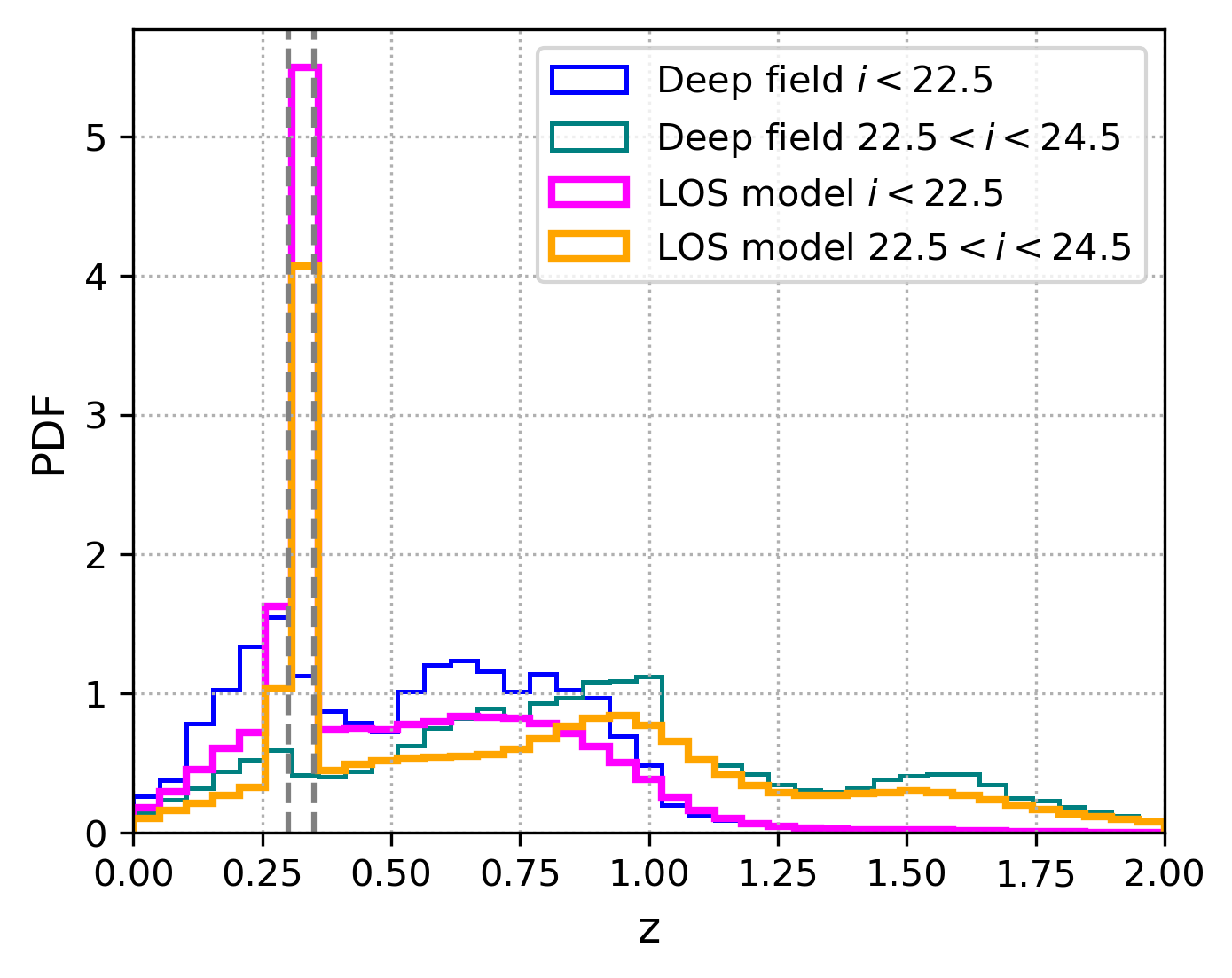}
    \caption{Line-of-sight model for the redshift distribution of galaxies near clusters with  $\lambda\in[45;\,60)$ and $z\in[0.3;\,0.35)$ within the projected radial range $R\in[1;\,3.16)$. \emph{Magenta, orange:} redshift distribution model around clusters in different magnitude bins. \emph{Blue, green:} photometric redshift distribution measured in the DES Deep Fields in different magnitude bins. \emph{Grey dashed:} limits of the cluster redshift range.  The cluster line-of-sight models show a significant deviation from the field line-of-sight, concentrated in a narrow redshift peak at $z_{\rm clust}$. }
    \label{fig:los_pz}
\end{figure}

The result of the statistical model is a set of random samples drawn from the feature PDF of the extrapolated cluster member galaxies, and a set of random samples which are drawn from the extrapolated field galaxy population. For both of these samples a KDE is constructed according to \autoref{sec:kde}, whose purpose is to provide a computationally efficient way of generating further samples. This model covers the full set of features listed in \autoref{tab:feature_list} to a deeper limiting magnitude of $i=24$ and is shown on \autoref{fig:joint model} for a single cluster bin with $\lambda\in[45;\, 60)$ and $z\in[0.3;\,0.35)$. In the following we overview the noteworthy features reproduced by this model and present the line-of-sight structure and galaxy surface density distribution of our synthetic clusters. 

\subsubsection{Line-of-Sight Model}
\label{sec:results_los}

Our galaxy redshift distribution model  used for creating synthetic cluster lines-of-sight is illustrated  on \autoref{fig:los_pz} for a cluster sample with $\lambda\in[45;\,60)$ and $z\in[0.3;\,0.35)$ where the emulated redshift PDF of galaxies with $i < 22.5$ and within the radial range $R\in[1;\,3.16)$ arcmin is shown as the magenta histogram. This is a combination of a cluster member term located at the mean cluster redshift $z=0.325$, and a field term. As a comparison the redshift PDF of deep-field galaxies is shown in blue for the same magnitude range. Owing to the extrapolation part of the analysis, the reconstructed line-of-sight is modeled down to the deep-field limiting magnitude of $i<24.5$.  It contains a faint cluster member population in addition to the faint end of the field galaxy population shown as the orange histogram, with the comparison redshift distribution of the deep-field galaxies shown as the green histogram.

This line-of-sight model incorporates galaxy redshifts derived from the deep-fields using $ugrizJHK$ bands. In turn the reduced redshift uncertainty for deep-field galaxies allows us to take the lens geometry correctly into account to apply the lensing effect for each galaxy. \autoref{fig:los_pz} also shows that the redshift distribution of galaxies near a cluster in projection is significantly different from the one in the Deep Fields. This aspect of the line-of-sight model enables us to construct mock observations where we can test the response of photometric redshift estimates to the presence of the galaxy cluster. This manifests itself as the problem of \emph{boost factors} or cluster member contamination \citep{Sheldon04.1, rmsva, Varga2018}, as well as propagating blending-related photometry effects onto the performance estimates of photometric redshifts.

\begin{figure*}
    \centering
    \includegraphics[width=\linewidth]{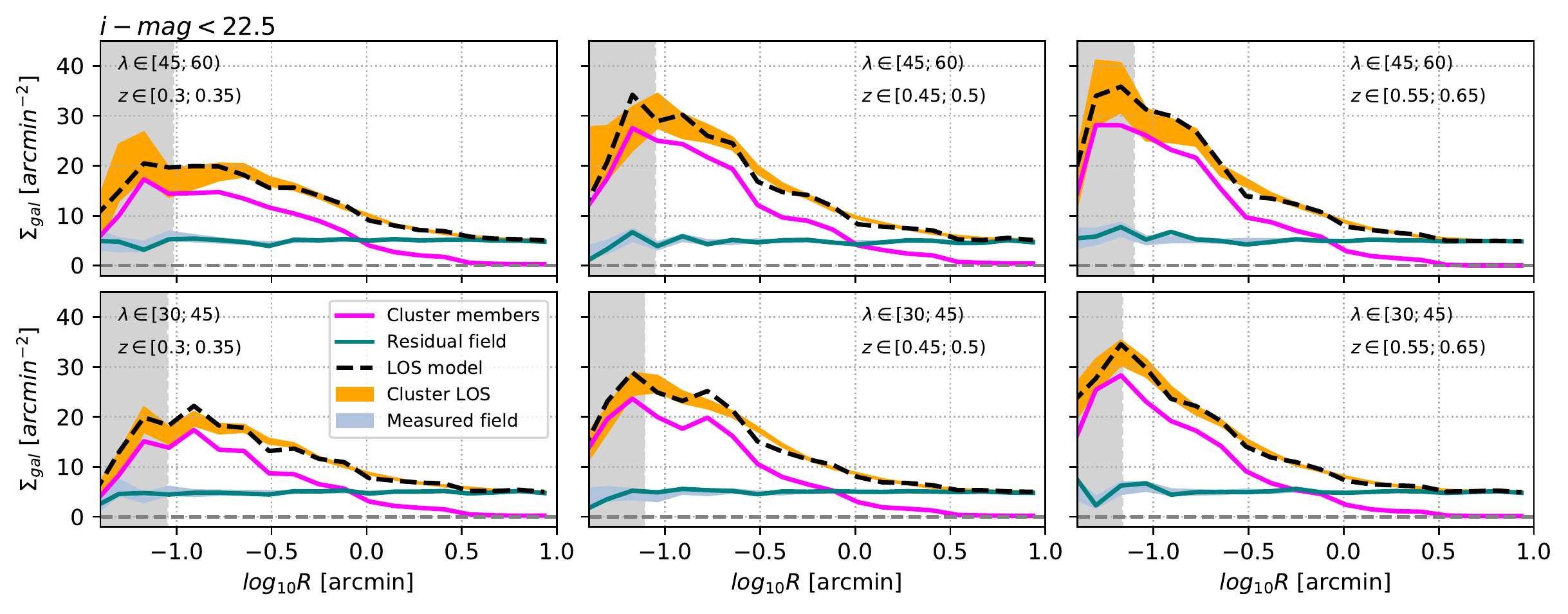}
    \caption{Surface density of galaxies around galaxy clusters with different richness and redshift. \emph{Orange:} Surface density profile measured around \redmapper\ clusters. The width of the shaded area represents the Poisson uncertainty propagated into surface density. \emph{Gray vertical area:} effective size of the cluster BCG ($\sqrt{T}$). The drop of the cluster LOS profile within this range represents a detection incompleteness due to the light of the central galaxy. In our model this regime is instead described by the BCG + ICL component (see \autoref{sec:icl}, compare with \autoref{fig:icl_example}).
    \emph{Gray:} Surface density of galaxies measured around reference random points. \emph{Green:} model for the surface density profile of  \emph{field} galaxies within the cluster line-of-sight. \emph{Magenta:} model for the surface density profile of \emph{cluster member} galaxies in the cluster line-of-sight. \emph{Black dashed:} model for the total galaxy surface density profile in the cluster line-of-sight (the sum of the green and magenta curves).
    }
    \label{fig:los_grid}
\end{figure*}

\subsubsection{Surface Density Model}
The models for the galaxy surface density profiles  are shown on \autoref{fig:los_grid}. The magnitude range is restricted to $i<22.5$. In addition, the measured galaxy surface density profile is indicated by the orange shaded area, and the surface density profile around the corresponding sample of reference random points as the gray shaded area. The width of these areas indicates the Poisson uncertainty of the number of galaxies.

The model for the field population is shown as the green lines  on  \autoref{fig:los_grid}. This distribution corresponds to the background model during the statistical background subtraction, but it is constructed by re-weighting and resampling deep-field galaxies. The excellent agreement between this and the profile measured around random points in the DES wide-field data is a strong consistency test of the statistical model, and is an indication that the  statistical background subtraction works as intended. 

The model for the pure cluster member distribution is shown as the magenta curves on \autoref{fig:los_grid}, and it captures the radial variations in surface density,  approaching zero at large radii, consistent with the finite extent of the cluster galaxy populations.  
The model for the full surface density profile is then obtained as the sum of the cluster member (magenta) and the field (green) population estimates, and this surface density profile is shown as the black dashed lines, which can then be directly compared with the galaxy profiles measured in the DES data around clusters (orange lines). The two show excellent agreement. The downturn of the surface density profiles at $R<0.1$ arcmin is due detection incompleteness caused by the central galaxy. In our model this regime is however described by the BCG + ICL component components (see \autoref{sec:icl}, compare with \autoref{fig:icl_example}). The light profile of cluster centrals do show considerable variability on such small scales \citep[see Fig. 18.][]{Kluge}, this is however not incorporated in the smooth ICL model of \citet{Gruen2018_ICL} adopted in this study.



\begin{figure*}
    \centering
    \includegraphics[width=\linewidth]{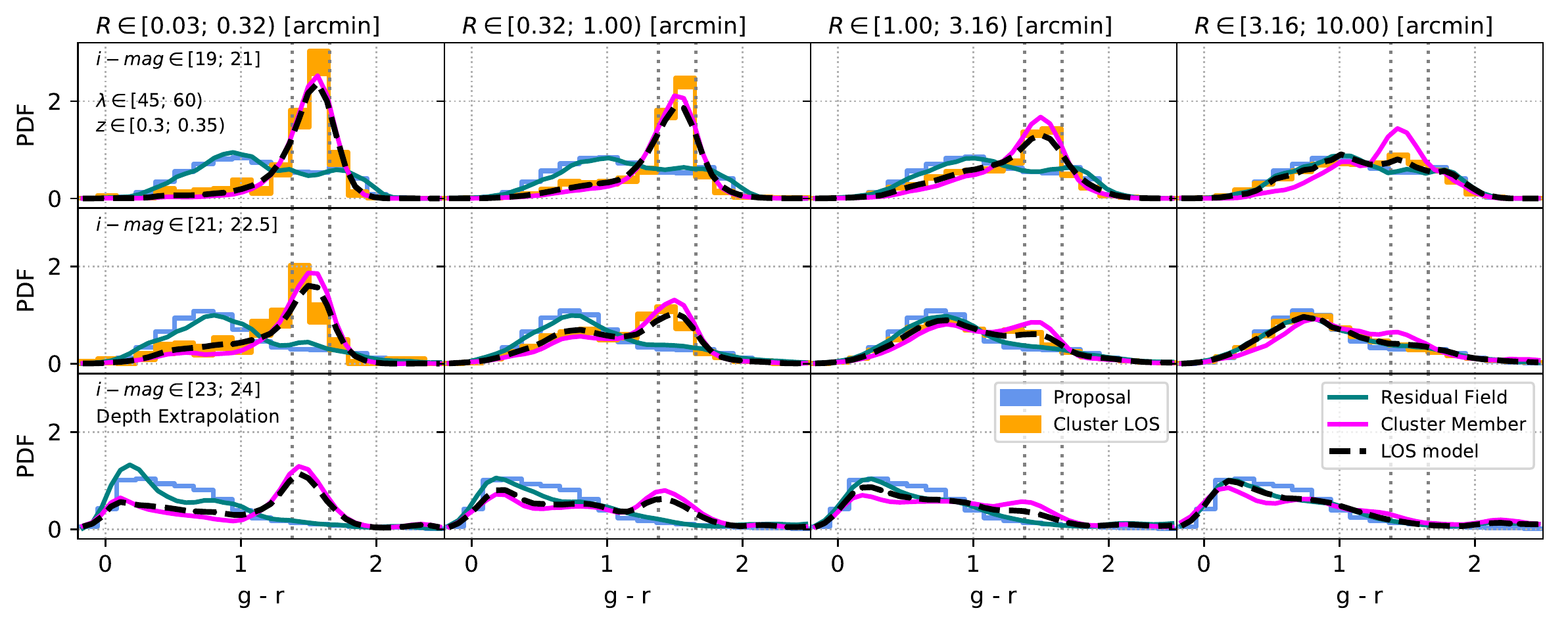}
    \caption[Conditional color distribution of galaxies around galaxy clusters]{Conditional color distribution of galaxies around galaxy clusters across four projected radial regimes (shown in the different columns) around galaxy clusters with $\lambda\in[45;\,60)$ and $z\in[0.3;\,0.35)$. The distribution of galaxies are  shown in $g-r$, $g-r$ and $r-i$ colors respectively. There are three magnitude ranges shown (rows), the first two $[19;\,21)$ and $[21;\,22.5)$ are fitted to the DES wide-field data, while the third $[23;\,24)$ is a pure extrapolation based on the algorithm. 
    \emph{Orange}:  color PDF measured as a histogram around galaxy clusters in DES data. The height of the shaded area indicates the Poisson uncertainty propagated into the normalized histogram.
    \emph{Blue}: color distribution measured within the corresponding magnitude range in the DES Deep Fields. This distribution is identical for each column and for all cluster samples.
    \emph{Green}: Model for the color distribution of  foreground and background galaxies in the line-of-sight.
    \emph{Magenta}: Model for the color distribution of cluster member galaxies.
    \emph{Black dashed}: Model for the full line-of-sight, which can be directly compared with the orange histogram.
    \emph{Gray dotted}: $1\sigma$ location of the \redmapper\ red-sequence cluster member galaxies.
    }
    \label{fig:color_mag_panel1}
\end{figure*}

\subsubsection{Cluster Member and Field Galaxy Features}

Galaxy clusters host a characteristic population of quiescent red galaxies distributed along the red-sequence, and also a non-red cluster member component. In projection, these cluster members are mixed together with foreground and background galaxies. 

\autoref{fig:color_mag_panel1} shows the model and  measurements for the $g-r$ color distribution of galaxies as an illustration of the statistical learning model for the cluster sample with  $\lambda\in[45;\,60)$ $z\in[0.3;\,0.35)$. The columns correspond to different bins of projected radius, and the rows to different magnitude ranges. The first two $[19;\,21)$ and $[21;\,22.5)$ rows show the model fitted to the DES wide-field data, while the third $[23;\,24)$ is a pure extrapolation based on the algorithm.
The measured color distributions from the DES wide-field data are shown as the orange histograms, with the colored area representing the Poisson uncertainty of the measurement. As a comparison, for each cell the respective conditional color distribution measured in the DES Deep Fields is shown (blue histogram). This population naturally has no radial dependence, and is thus identical in the different columns. 

Out of the above two populations, only the  deep-field one is measured down to the third magnitude bin $i\in[23;\,24)$, therefore the cluster measurement (orange) is not shown there. The color distribution around clusters shows a strong radial trend, with the orange histogram approaching the blue with increasing radius. A dominant driver of this trend is increasing prominence of the red-sequence at low radii, which manifests as a peak in the color distribution. The relative weight of the red-sequence is greater for brighter galaxies, and the difference between cluster and field lines-of-sight is also greater for brighter galaxies. As a reference, the  location of the \redmapper\ red-sequence model is indicated by the vertical gray dotted lines. These lines correspond to the $1\sigma$ range of the membership probability weighted color distribution of \redmapper\ cluster members for that cluster richness, redshift range. Both the location and the width of the peak of the cluster member  histogram (shown in orange) is consistent with the properties of \redmapper\ cluster members, indicating that it is indeed an imprint of the red-sequence.  We note that only galaxies with $L > 0.2\,L_\star$ are considered by \redmapper\ as  potential member galaxies and this does not fully cover the faintest magnitude bin of this analysis.


\autoref{fig:color_mag_panel1} shows the model for the projected galaxy distributions around galaxy clusters as the black dashed lines, which can be directly compared with the orange histogram. This model is derived without direct information about the wide-field galaxy luminosity function around clusters, and only using information from the deep-field data. Nevertheless, as visible on the upper two rows of \autoref{fig:color_mag_panel1}, the line-of-sight model can describe the magnitude dependent color variations of the galaxy distributions, and well approximate the relative weight of the red-sequence peak, albeit slightly over-estimating its width. The bottom row shows the model for galaxies in the line-of-sight with $i\in[23;\,24)$. Due to the extrapolation part of the approach, the model extends to these fainter magnitudes, even though they are not directly measured in cluster lines-of-sight.

The feature distributions of foreground and background galaxies are independent of the cluster galaxy population. Thus it is expected that the residual field model is independent of radius. While the bright tip of the DES Deep Fields is not fully representative of the actual median DES wide-field survey due to sample variance, it still provides a reasonable reference distribution. Comparing the residual field model (green curve) with the deep-field distribution (blue histogram) on \autoref{fig:color_mag_panel1} shows no strong radial variations. The residual field indeed approximates the deep-field distribution, with only minor deviations visible at the faint end.

\begin{figure}
    \centering
    \includegraphics[width=\linewidth]{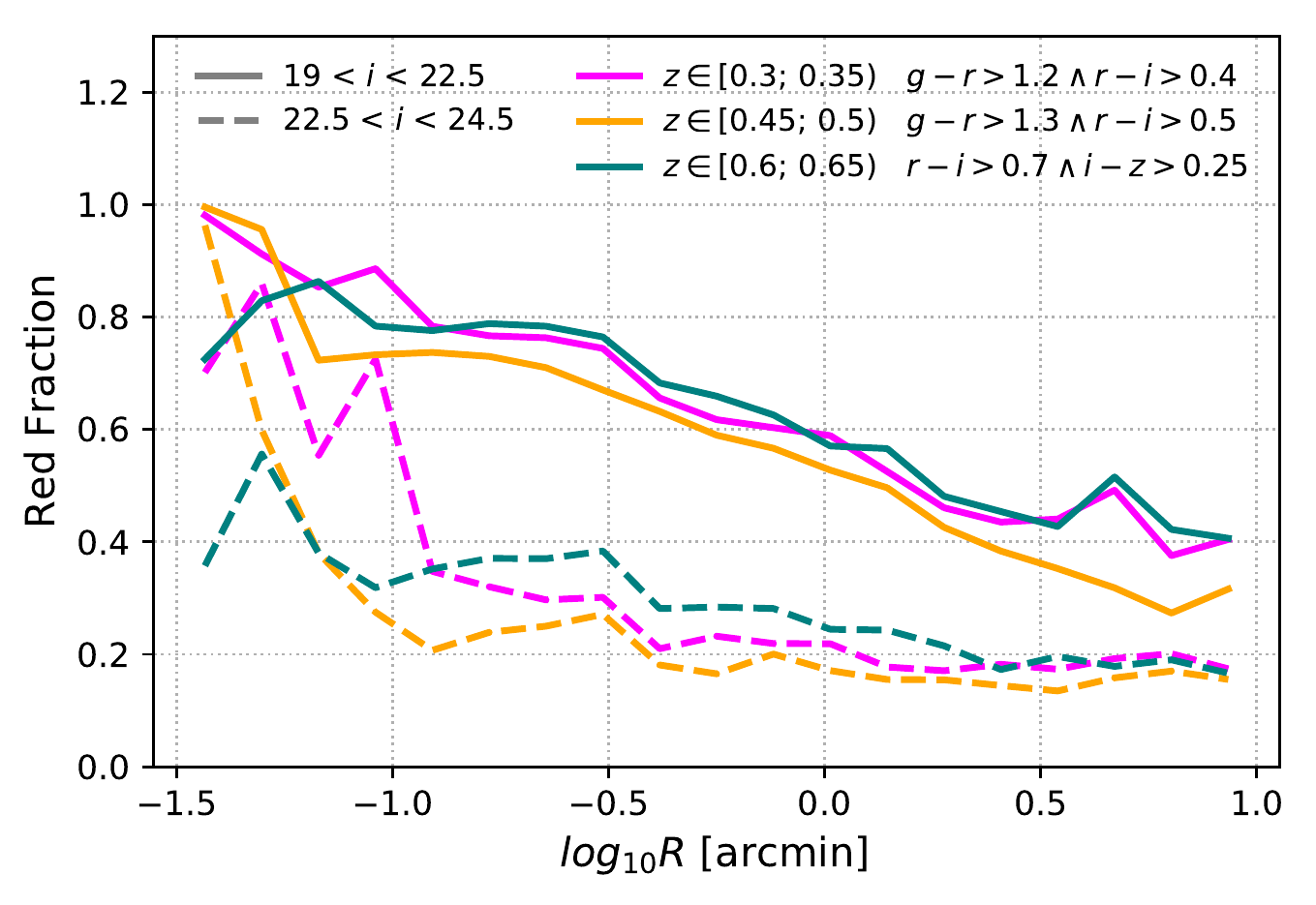}
    \caption{Red fraction of cluster members as a function of projected radius for three different cluster redshift samples with $\lambda\in[45;\,60)$.}\label{fig:los_redfrac}
\end{figure}

\subsubsection{Red Fraction Estimates}

The radial color evolution of the cluster member galaxy population can be described by the approximate red-fraction, whose radial profile for the three high richness bins is shown on \autoref{fig:los_redfrac}, along with the color cuts used in the definition. These regions are chosen to bracket the position of the red sequence which is dominant at low radii. Two magnitude ranges are shown: a brighter bin covering $i\in[19;\,22.5)$  coincides with the DES wide-field depth, and a fainter bin covering $i\in[22.4;\,24.5)$, which is derived from a purely extrapolated color-color distributions.  While the figure shows only the higher richness samples, there appears to be no significant difference between the richness bins. 

The bright galaxy sample shows a clear monotonic trend in all redshift and richness samples, where the  red-fraction decreases from approximately unity at very low projected radii to approximately 30 - 40 per cent at large radii approaching 10 arcmin. This behaviour is consistent with previous measurements \citep{Butcher78, Hansen08, Hennig2016}. It is also in agreement with existing DES-like synthetic clusters derived from decorated gravity-only numerical simulations presented in \citet{DeRose2018, Varga2018}. The same behaviour is not uniformly true for the fainter, extrapolated red-fraction profiles. Some cluster bins show a prominent red galaxy population at the center, the decline is much faster for these fainter populations than the brighter counterparts for the same clusters. At large radii the galaxy population appears to show a constant mix of red and blue members, and approach the preferentially bluer cosmic mean galaxy populations.

\section{Synthetic Observations}
\label{sec:random_los}

\subsection{Random Draws of Galaxy Populations}
\label{sec:draws}
The model for non-central galaxies is composed of two main components: the distribution of cluster member galaxies (satellites) and the distribution of foreground and background galaxies.  A synthetic cluster line-of-sight is created  by random draws from the PDF of the different components. Here each draw corresponds to adding a new galaxy to a mock catalog with an angular and redshift position, and the photometric and morphological features contained within the model. 

A PDF carries no information about the absolute number of objects, therefore this needs to be set based on the observed number of galaxies. In real observations only the bright end of the luminosity function is observed in the survey (i.e. $i<22.5$) therefore  the number of fainter galaxies must be defined according to their relative probability in the model.

A single mock galaxy cluster is constructed the following way:
\begin{enumerate}
\item For each radial range $l$, calculate $\hat{N}_{C;l}$ and $\hat{N}_{R;l}$ the mean number of galaxies with $i<22.5$ around clusters and random points respectively in radial range $l$.

\item For each radial range $l$, take a Poisson random number of galaxies based on the mean number as
\begin{equation}
N_{M;l} = \mathrm{Poisson}\left(\frac{\hat{N}_{C;l} - \hat{N}_{R;l}}{p_{\mathrm{memb};l}\left(i < 22.5 \right)} \right)\,,
\end{equation}
and
\begin{equation}
N_{R;l} = \mathrm{Poisson}\left(\frac{\hat{N}_{R;l}}{p_{\mathrm{rand};l}\left(i < 22.5 \right)} \right)\,.
\end{equation}

\item Draw cluster members $N_{M;l}$ times from $p_{\mathrm{memb};l}$ and foreground and background galaxies $N_{R;l}$ times from $p_{\mathrm{rand};l}$.

\item For cluster members set the redshift to $z_\mathrm{clust}$.

\item Convert the projected radius feature $R_i$ into 2D position assuming circular symmetry in a flat-sky approximation.

\end{enumerate}
The outcome of the above recipe is a galaxy catalog which contains cluster members and foreground and background galaxies each distributed according to their respective statistical models derived from the survey data, but extrapolated to a fainter limiting magnitude, and the surface density of galaxies is set to the mean surface density measured around galaxy clusters.

In practice we update step 1 by only measuring  $\hat{N}_{C;l}$ from data, and expressing $\hat{N}_{R;l}$ as a function of $\hat{N}_{C;l}$ using the statistical model. In practice this is achieved by taking the ratio of accepted events during the rejection sampling (see \autoref{sec:resampling})  which only fulfill \autoref{eq:rs_pfield_resampling}, to the amount of events which fulfill both \autoref{eq:rs_pfield_resampling} and \autoref{eq:rs_pmemb_resampling2}. This latter formulation avoids scenarios when due to measurement noise by chance $\hat{N}_{R;l} > \hat{N}_{C;l}$.

\subsection{Cluster Lens Model and Galaxy Shapes}
\label{sec:shear}

Synthetic weak lensing measurements require  a mass model for the galaxy cluster to  apply gravitational shear to the background galaxies. For this we make use of the mass models and mass constraints found in \citet*{rmy1}. As that analysis did not find a significant redshift evolution in the richness-mass scaling, we can approximate the relevant mean cluster masses for the present mocks, that is $M_\mathrm{200m}\approx 10^{14.45}\;M_\odot$ for the $\lambda\in[30;\,45)$ bin and $M_\mathrm{200m}\approx 10^{14.65}\;M_\odot$ for the $\lambda\in[45;\,60)$ bin across the three different redshift bins. 

In the following pathfinder study, we only consider the mass model for the 1-halo term which is dominant on the small scales explored in this study, and consists of a spherically symmetric mass distribution with Navarro-Frenk-White (NFW) mass profile  \citep{Navarro96.1}. This lens mass distribution is placed at the cluster redshift $z_\mathrm{clust}$ and subsequently gravitational shear and magnification is applied to line-of-sight galaxies based on their true redshifts assigned by the model. The lensing effect induced by a NFW halo is expressed analytically following \citep{Wright}. Reduced gravitational shear $\mathbf{g}$ is directly applied to each galaxy through the \texttt{ngmix} \texttt{bdf} galaxy model. The magnification ($\mu$) is however only applied as a simple approximation, by modulating the total flux of the galaxy light models $F_{\mathrm{lensed};\,i} = \mu_i\; F_{i}$ in an a-chromatic way. This correctly captures the change in the total observed flux of each galaxy, but does not reproduce the increase in observed size. The impact of this approximation is expected to be minor given the very small apparent size of the high-redshift galaxies which experience the greatest magnification effect.

\subsection{BCG and Intra-Cluster Light Model}
\label{sec:icl}

 \begin{figure}
    \centering
    \includegraphics[width=\linewidth]{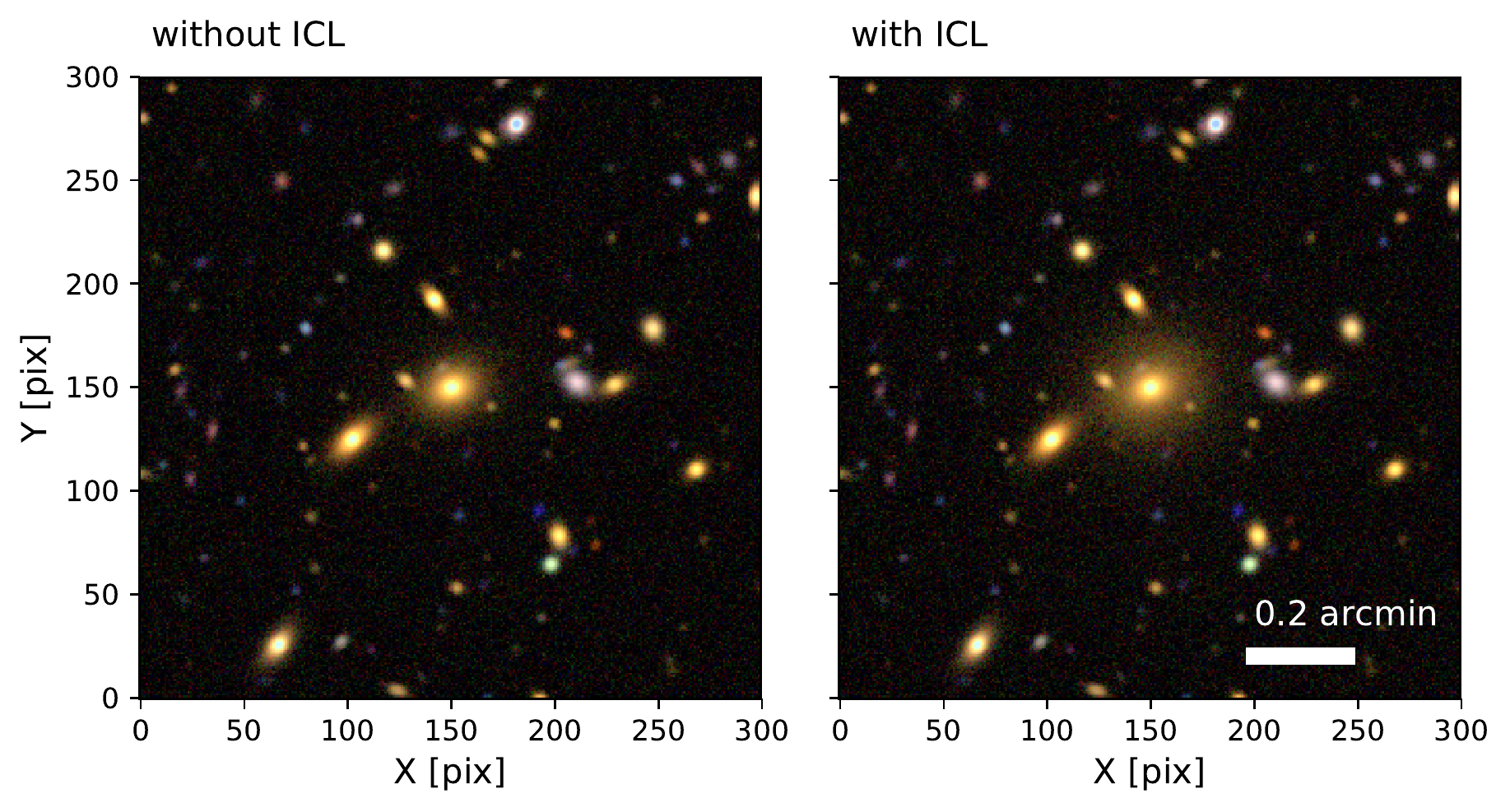}
    \caption{Synthetic center of a mock galaxy cluster without (left) and with the intra-cluster light model applied (right). Real galaxy clusters host a large fraction of their stellar light in the form of ICL, which the simple BCG only light model cannot reproduce. This is seen on \autoref{fig:cluster_schematic} and \autoref{fig:mock_cluster_composit}.}
    \label{fig:icl_example}
\end{figure}

A prominent feature of galaxy clusters is the presence of a \emph{bright central galaxy} (BCG) and a surrounding distribution of \emph{intra-cluster light} (ICL) emitted by a diffuse stellar component bound to the cluster halo. These components contain a significant fraction of the total optical light emitted by the cluster  \citep{Zhang19, Santos2020, Kluge}, therefore accounting for them is essential in a dedicated simulation of synthetic galaxy cluster observations.

By construction galaxy clusters identified by \redmapper\ are always centered on a bright red-sequence galaxy. This is a simplified view of reality, as in recent mergers or in non-equilibrium systems the central galaxy might not be red or the brightest, or there might be multiple similarly bright BCGs \citep{Rykoff2014_RM1}. Originating from the special location they inhabit, the central galaxies of massive halos follow a different evolutionary track compared to satellite galaxies. It is observed that their properties are closely tied to the mass and properties of their cluster \citep{Postman95}, and their luminosity function is approximately Gaussian at fixed cluster mass proxy and redshift \citep{Hansen08}.
Based on these observations we model the synthetic central galaxy in the mocks as having the mean properties of the \redmapper\ central galaxies in the cluster sample. The relevant mean central galaxy features are listed in  \autoref{tab:bcgs} for the different cluster redshift and richness samples. The central galaxies are assumed to have a de Vaucouleurs light profile, and the only stochastic element in the model is their random orientation in the plane of the sky with fixed ellipticity $|g|$. 
 
The total light in the central region of a cluster is, however, not fully described by the above model, as there is a continuous transition between the light usually associated with the central galaxy and the intra-cluster light \citep{Kluge}. \citet{Zhang19} investigated the properties of the ICL for \redmapper\ selected galaxy clusters with $z_\mathrm{clust}\in[0.2;\,0.3)$ within the DES Y1 dataset. In a stacked analysis they measured the diffuse light of the ICL down to a surface brightness of 30 $\mathrm{mag} \,\mathrm{arcsec}^{-2}$. \citet{Zhang19} investigated the richness (mass) dependence of the ICL, finding a self-similarity of the light profile when expressed in units of $R_{200m}$. The ICL - mass relation was further established by \citet{Santos2020} in an expanded re-analysis of the DES Y1 \redmapper\ cluster sample.  Using the measurements of \citet{Zhang19}, \citet{Gruen2018_ICL} constructed a simple model for the ICL observed around \redmapper\ clusters in DES. This model extrapolates from the measurement of \citet{Zhang19} in terms of cluster mass using the self-similarity of the profiles, and also in terms of cluster redshift by assuming a simple passively evolving stellar population within the ICL. We note that this latter assumption is closely related to the formation history and age of the ICL, which is poorly constrained from current observational studies due to the difficulty of high redshift observations. Thus in case of a late-forming ICL the above extrapolation overestimates the total light contained in it at early times. Furthermore, the model neglects the mild radius dependent color gradient in the ICL, where the outer ranges are slightly bluer.
 
In the following we adopt the ICL model of \citet{Gruen2018_ICL}. As a simplification we assume that the colors of the ICL are identical to the mean colors of BCGs at that redshift and cluster richness sample. The ICL component extends to large radii as an approximate power law surface density light profile, while the \texttt{ngmix} BCG light model is dominant in the inner regions. Because of their overlap, these components cannot be directly added to each other. Therefore we define a tapered ICL model where the tapering scale is set by the size of the BCG component $\theta_S = \sqrt{T_{BCG}}$, where $T_{BCG}$ is taken from the DES Y3 MOF photometry catalog and is defined the same way as the size parameter listed in \autoref{tab:feature_list}. To ensure the smooth joining of the BCG and ICL components we define the total light profile model as
\begin{equation}
\mu(\theta) = \mu_{BCG}(\theta) + \left(1 - \frac{1}{1 + e^{2\,(\theta - \theta_S)}} \right)\mu_{ICL}(\theta)\,.
\end{equation}
An illustration of this joint BCG + ICL light profile in the mock cluster images is shown on \autoref{fig:icl_example}. The two panels show an identical set of mock galaxies for a synthetic cluster corresponding to the cluster bin with $\lambda\in[45;\,60)$ and $z\in[0.3;\,0.35)$,  however the left panel shows only the \texttt{ngmix} galaxy models, while the right panel also shows the ICL component added.

\subsection{Survey-like Images}
\label{sec:galsim}
Simulated galaxy images are the bedrock of estimating the performance of weak lensing methods, and therefore they were the topic of extensive study in the literature (\citealp{STEP2,GREAT08,GREAT3, Jarvis2016}, \citealp*{Y1shape}, \citealp{Samuroff18}). In the following we make use of a simplified version of the image simulation pipeline developed for the Y3 analysis of DES \citep{Y3sim}.

The construction starts with a catalog of photometric objects which will inhabit the mock image. For this study this catalog contains the parameters of the \texttt{ngmix} \texttt{bdf} light distribution model for each entry which are pixel position in the image, shape $(g_1;\,g_2)$, size $T$, bulge / disk flux fraction, and fluxes in  $g,r,i,z$ bands. This catalog corresponds to a random realization of a mock line-of-sight constructed according to \autoref{sec:draws} and \autoref{sec:shear}. Finally the central galaxy is added as defined in \autoref{sec:icl}.
At this stage stars and foreground objects can be added according to their density at the targeted galactic latitude.  In the present pathfinder study these are drawn from the population of stars excluded in \autoref{sec:indexer}. Furthermore,
we only consider a simplified scenario and add a stellar sample drawn from the deep-field catalog according to their relative density in the deep-field footprints. 

Synthetic images are created via a customized version of the DES Y3 image simulation pipeline \citep{Y3sim}, which renders images based on a galaxy image simulation package \texttt{GalSim} \citep{Rowe2014}, while using an extension package for the  \texttt{ngmix} \texttt{bdf} light profile model used in the actual DES Y3 deep-field analysis\footnote{\url{https://github.com/esheldon/ngmix}, the \texttt{ngmix.gmix.GmixBDF} model.}. This model describes the galaxies as a combination of two terms: an exponential light profile (disk) and a de Vaucouleurs (bulge) light profile. Given that most galaxies in a DES-like survey are poorly resolved, an additional constraint is enforced by setting the effective radius of both light profile components to be identical. 

In the following, we consider a simplified setup of the observational scenario of DES where we directly simulate the so-called \emph{co-added} survey images. Under real circumstances due to variations in observing conditions and the point spread function (PSF) between exposures the net PSF in co-added images is difficult to model, thus the DES shape estimation pipeline itself takes single exposure images as input. In a simulation such  variations can be factored out, which allows us to simplify the simulation setup into deeper mock co-added images with well behaved PSFs.

The synthetic co-added images are constructed the following way:
\begin{enumerate}
\item The image canvas is defined with its desired dimensions and pixel scale, in the case of DES, 0.27 arcsec / pixel. The canvas is defined as a 10k$\times$10k pixel rectangle. 

\item For each object a small cutout image (postage stamp) is constructed. The light model is defined using \texttt{ngmix}, convolved with a representation of the mock PSF, then rendered into a postage stamp. We model the PSF as a Gaussian with a full-width half-maximum (FWHM) of 0.9 arcsec, which is roughly equal to the median DES observing condition \citep{Y3gold}.

\item After the creation of all postage stamps, they are added onto the main canvas at their intended pixel positions.

\item A noise map is applied to the image. In  this study we take the noise properties of a randomly selected DES tile (DES2122+0209) and apply Gaussian noise matched to reproduce the median flux of the unmasked regions of the reference tile in the chosen observational band.  Choosing the noise level for synthetic images is not straightforward, as a substantial amount of light which is traditionally attributed to noise in fact originates from undetected faint stars and galaxies \citep{Hoekstra17, Euclid19, Eckert20}. In the framework of the present analysis many of these undetected sources are explicitly part of the rendered objects, therefore as a rough approximation we reduce the background noise variance by half for illustration purposes. 

\item Finally the tapered ICL model defined according to \autoref{sec:icl} is evaluated for the pixel positions of the mock image and the additional light component  is added onto the synthetic observation. We assume that the ICL has the same ellipticity and major axis direction alignment as the central galaxy.
\end{enumerate}

The result of this recipe is illustrated on \autoref{fig:cluster_schematic} where a $gri$-band color composite image is shown for synthetic clusters side by side with \redmapper\ clusters with similar observable parameters. While the synthetic images do contain an approximate stellar population based on faint stars observed in the Deep Fields, very bright stars which need to be masked are not currently reproduced in the mock observations. Furthermore, low redshift foreground objects such as galaxies with visible disc and spiral arm features are not contained in the scope of the present analysis.
In addition to the color composite images, \autoref{fig:cluster_schematic} also illustrates the composition of the lines-of-sight. The third row of each figure shows the brightness distribution of the cluster component with brown/red symbols, and the foreground and background component with blue symbols. The shade and size of the symbols indicate the brightness with fainter objects shown as smaller markers. Many of the faint objects are barely or not at all discernible on the composite images. Yet these unresolved sources influence the performance of photometric methods (\citealp{Hoekstra17, Euclid19}, \citealp*{Y3Balrog}). The bottom row of each figure shows the exaggerated gravitational shear imprinted on background sources (the ellipticities are increased by a factor of 20). The background sources are shown in as darker color for low redshift and lighter color for high redshifts. Cluster members are shown in black symbols, while foreground objects are shown in green. The different brightness values are indicated by the different marker sizes.

\begin{figure*}
    \centering
    \includegraphics[width=1.\linewidth]{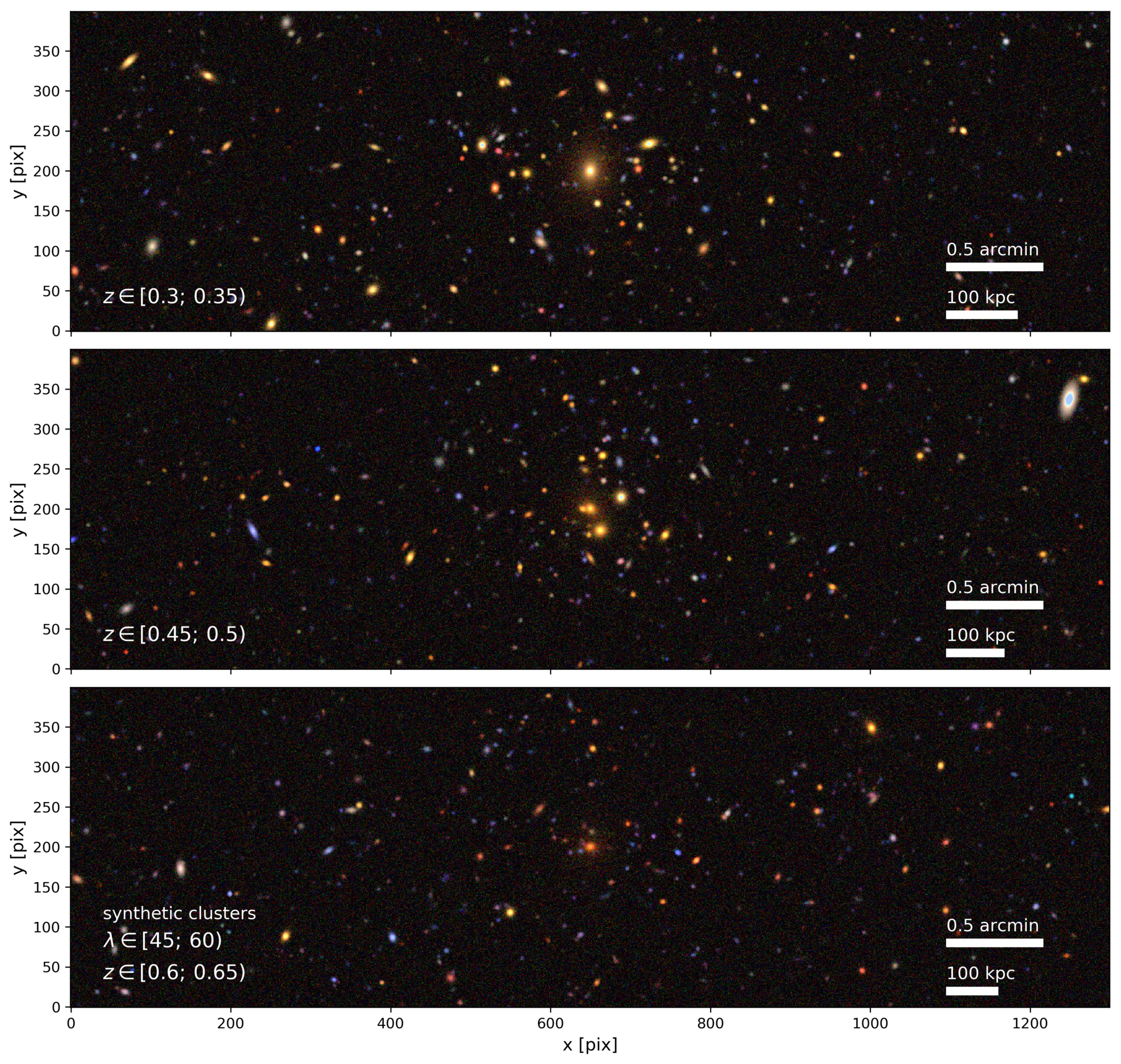}
    \caption[Synthetic galaxy clusters corresponding to \redmapper\ clusters with $\lambda\in[45;\;60)$ across the different redshift ranges]{Synthetic galaxy clusters corresponding to \redmapper\ clusters with $\lambda\in[45;\;60)$ across the different redshift ranges.}
    \label{fig:mock_cluster_composit}
\end{figure*}

While the galaxy populations of the $\lambda\in[30;\,45)$ and $\lambda\in[45;\,60)$ bins are found to be  close in terms of their galaxy surface density profiles, clusters show greater differences between the different redshift ranges. This is illustrated by \autoref{fig:mock_cluster_composit}, which shows synthetic galaxy clusters with $\lambda\in[45;\,60)$ in the $z\in[0.3;\,0.35)$, $z\in[0.45;\,0.5)$ and $z\in[0.6;\,0.65)$ cluster samples. These color composite images show a striking illustration of the changes in the visible properties of galaxy clusters across cosmic time.

\section{Summary and Conclusions}
\subsection{Method Overview}

We present a pathfinder study to generate synthetic galaxy clusters and cluster observations in an unsupervised way from a combination of observational data taken by the Dark Energy Survey up to its third year of observations (DES Y3). Example realisations of synthetic galaxy cluster observations are shown on \autoref{fig:cluster_schematic} and \autoref{fig:mock_cluster_composit}.  Galaxy clusters present a unique challenge for validating weak lensing measurements due to the increased blending among light sources, the presence of the  intra-cluster light (ICL), and the characteristically stronger shear imprinted on source galaxies. The aim of these synthetic observations is to enable future studies to address the above factors by calibrating and validating the performance of galaxy cluster weak lensing in an end-to-end fashion from photometry, through shear and photometric redshift measurement and calibration to mass recovery from lensing profiles or lensing maps in a fully controlled environment. 
The focus of this paper is to introduce the statistical learning algorithm itself and to demonstrate a pilot implementation for DES Y3 data. This consists of the following steps:
\begin{itemize}
\item We measure the galaxy content of \redmapper\ galaxy clusters and their sky environments in projection, as a function of cluster richness and redshift (\autoref{sec:indexer}).

\item Develop and validate a KDE framework for  representing galaxy distributions as high-dimensional probability density functions of photometric and morphological features \autoref{sec:kde}). This KDE generalizes the finite set of galaxy and cluster observations into a continuous model, and provides a numerically efficient, extendable framework for accommodating potential new galaxy features from external data.

\item Derive a mathematical formalism to combine wide-field and deep-field survey data, augmenting and extrapolating our model beyond the depth and scope of the wide-field data (\autoref{sec:extrapolation}).

\item Create a model for the cluster member galaxy content of \redmapper\ clusters via statistical background subtraction in a multi-dimensional feature space (\autoref{sec:resampling}).

\item Through a series of comparisons between the properties of observed and modeled galaxies drawn from the KDE, we demonstrate an excellent agreement in terms of real and synthetic galaxy catalogs of cluster lines-of-sight (\autoref{sec:model_results}). We note that this reflects primarily on the performance of the input catalogs used in creating the synthetic observations. A detailed analysis of the agreement between real data and the photometry derived from the  synthetic images is delegated for future work. Corrections for the potential incompleteness of synthetic images can be  addressed as a prior for \autoref{eq:p_formula}.


\item Combine the above steps into an algorithm constructing and rendering new realizations of mock galaxy clusters into synthetic images (\autoref{sec:random_los}).
\end{itemize}

This work addresses four distinct problems arising with simulated data:
\begin{enumerate}[label=\Alph*]

\item The method does not rely on numerical simulations of baryonic structure formation and galaxy evolution to construct galaxy clusters and thus it is independent from assumptions and approximations inherent in cosmological simulations.

\item Synthetic galaxy clusters are generated to match their observed galaxy content in DES Y3. Extrapolations of the galaxy populations are performed where necessary, based  on observational data.

\item The algorithm is formulated as a transparent, explicit recipe. Therefore the different components can be readily modified where necessary and external information (e.g. survey incompleteness corrections, priors on cluster galaxy properties) can be added in a principled way.

\item Via the statistical learning approach, new, statistically independent realizations of synthetic galaxy cluster observations can be created at minimal computational cost. 

\end{enumerate}
Finally, the generative cluster galaxy model encapsulates the properties of cluster member galaxies in DES Y3 observations, and thus can be used as a validation or augmentation dataset for the results of  numerical galaxy cluster simulations.

\subsection{Future Outlook}

Due to the inherent complexity and scope of a full cluster weak lensing systematics control analysis, the overall effort is divided into multiple stages, of which this paper presents the initial step, and defines the framework for a data driven, customizable, generative cluster model. Upcoming studies will focus on integrating the synthetic cluster image generation into the weak lensing analysis pipeline of DES, and following that will perform a direct end-to-end  calibration for cluster lensing systematics. Since the synthetic cluster images mimic the observational setting of the real survey, applying standard survey data processing pipelines is expected to require only minor adaptations in analysis choices, and will provide the same data products as the real measurement. Of particular interest will be the quantification of detection efficiency in the crowded environments near cluster centers, and the impact of ICL and blending on the photometry solutions. These systematics propagate to photometric redshift errors, which we will be able to directly quantify.  Similarly, running shear measurement pipelines on the synthetic images will allow a direct measurement on any additive or multiplicative shear bias caused by the presence of the ICL and cluster member galaxies. The primary outcome of the above steps will be to quantify the scale dependent shear and photometric redshift bias induced by galaxy clusters, as a function of their observable features (e.g. redshift, richness, or other mass proxy). Due to the modular nature of the recipe for generating galaxy clusters, various ingredients (e.g. ICL, cluster member morphology) can be turned off for parts of the analysis, allowing to also constrain their \emph{specific} impact on shear and photo-z bias. Such correction profiles are already used in literature to account for cluster member contamination, and can be propagated to the mass-observable during the likelihood analysis \citep*{rmy1}.

The planned analysis will be made possible in two distinct configurations: While the use-case described in this paper focuses on full line-of-sight image simulations, cluster-only images can also be straightforwardly generated to allow for mock image injections into the real survey observations in a manner similar to \cite*{Y3Balrog}.

A further future direction is increasing the realism and plausibility of the generative galaxy cluster model. The presented implementation aims to reproduce the stacked observational scenario, while using only those datasets available within DES. Nevertheless, our framework is designed to allow easy augmentation with external data, such as numerical cosmological simulations of galaxy clusters (e.g.  Magneticum, \citealp{Dolagprep}; or illustrisTNG, \citealp{Nelson2019}), while ensuring that the final cluster model remains consistent with observations. These augmentations would take the form of replacing the $p_\mathrm{prop}$ taken from the DES deep fields with the appropriate KDE model from the chosen external data. By using a proposal distribution already informed by the feature PDF of real cluster members, that information will be propagated to the  generative cluster model.

Deviations from the mean stacked line-of-sight model can be implemented by allowing the BCG and ICL properties, and the mass model to be also drawn from distributions, rather than being fixed to the mean value for each stack. Given a model for intrinsic or correlated scatter between BCG, member galaxy properties or the mass model, a further layer of rejection sampling can be added in \autoref{sec:random_los}. In that additional layer, from many realizations of galaxy cluster catalogs, subsets can be filtered out which reproduce the desired intrinsic or correlated scatter. Furthermore, in case there is access to a preferential direction in individual clusters (e.g. miscentering offset from multi-wavelength centroids, or cluster ellipticity major axis direction), that can be incorporated by replacing the scalar R in the formalism with a 2D relative position $\vec{R} = ( R_1, R_2)$, augmenting the default circular symmetry of the cluster model. The presented KDE framework is designed anticipating such extension features, therefore their incorporation to the generative cluster model is expected to be straightforward.


While this work was done in preparation of a cluster weak lensing analysis using the DES Y3 data, owing to the transparent and modular nature of the presented recipe it is expected that the algorithm can be fitted to other similar weak lensing surveys with minimal effort.
Given their great statistical power, current (DES, \citealp{DESproposal}; KiDS, \citealp{KiDS2013}; HSC, \citealp{HSC_Survey}) and upcoming (Rubin Observatory, \citealp{LSST2019}; Euclid, \citealp{Euclid2011}; Roman Space Telescope, \citealp{WFIRST2015}) weak lensing surveys are increasingly dominated by systematic uncertainties. For this reason, calibration and validation tools such as the one presented in this study will be indispensable in exploiting the cosmological and astrophysical information made accessible by large area sky surveys.

\section*{Data Availability}
The data underlying this article will be made available according to the data release schedule of the Dark Energy Survey.

\section*{Acknowledgments}
This research was supported by the Excellence Cluster ORIGINS which is funded by the Deutsche Forschungsgemeinschaft (DFG, German Research Foundation) under Germany's Excellence Strategy - EXC-2094-390783311. The calculations have been in part carried out on the computing facilities of the Computational Center for Particle and Astrophysics (C2PAP). This work was supported by the Department of Energy, Laboratory Directed Research and Development program at SLAC National Accelerator Laboratory, under contract DE-AC02-76SF00515 and as part of the Panofsky Fellowship awarded to DG.

Funding for the DES Projects has been provided by the U.S. Department of Energy, the U.S. National Science Foundation, the Ministry of Science and Education of Spain, the Science and Technology Facilities Council of the United Kingdom, the Higher Education Funding Council for England, the National Center for Supercomputing Applications at the University of Illinois at Urbana-Champaign, the Kavli Institute of Cosmological Physics at the University of Chicago, 
the Center for Cosmology and Astro-Particle Physics at the Ohio State University,
the Mitchell Institute for Fundamental Physics and Astronomy at Texas A\&M University, Financiadora de Estudos e Projetos, 
Funda{\c c}{\~a}o Carlos Chagas Filho de Amparo {\`a} Pesquisa do Estado do Rio de Janeiro, Conselho Nacional de Desenvolvimento Cient{\'i}fico e Tecnol{\'o}gico and 
the Minist{\'e}rio da Ci{\^e}ncia, Tecnologia e Inova{\c c}{\~a}o, the Deutsche Forschungsgemeinschaft and the Collaborating Institutions in the Dark Energy Survey. 

The Collaborating Institutions are Argonne National Laboratory, the University of California at Santa Cruz, the University of Cambridge, Centro de Investigaciones Energ{\'e}ticas, 
Medioambientales y Tecnol{\'o}gicas-Madrid, the University of Chicago, University College London, the DES-Brazil Consortium, the University of Edinburgh, 
the Eidgen{\"o}ssische Technische Hochschule (ETH) Z{\"u}rich, 
Fermi National Accelerator Laboratory, the University of Illinois at Urbana-Champaign, the Institut de Ci{\`e}ncies de l'Espai (IEEC/CSIC), 
the Institut de F{\'i}sica d'Altes Energies, Lawrence Berkeley National Laboratory, the Ludwig-Maximilians Universit{\"a}t M{\"u}nchen and the associated Excellence Cluster Universe, 
the University of Michigan, the National Optical Astronomy Observatory, the University of Nottingham, The Ohio State University, the University of Pennsylvania, the University of Portsmouth, 
SLAC National Accelerator Laboratory, Stanford University, the University of Sussex, Texas A\&M University, and the OzDES Membership Consortium.

Based in part on observations at Cerro Tololo Inter-American Observatory, National Optical Astronomy Observatory, which is operated by the Association of 
Universities for Research in Astronomy (AURA) under a cooperative agreement with the National Science Foundation.

The DES data management system is supported by the National Science Foundation under Grant Numbers AST-1138766 and AST-1536171.
The DES participants from Spanish institutions are partially supported by MINECO under grants AYA2015-71825, ESP2015-66861, FPA2015-68048, SEV-2016-0588, SEV-2016-0597, and MDM-2015-0509, 
some of which include ERDF funds from the European Union. IFAE is partially funded by the CERCA program of the Generalitat de Catalunya.
Research leading to these results has received funding from the European Research
Council under the European Union's Seventh Framework Program (FP7/2007-2013) including ERC grant agreements 240672, 291329, and 306478.
We  acknowledge support from the Australian Research Council Centre of Excellence for All-sky Astrophysics (CAASTRO), through project number CE110001020.

This manuscript has been authored by Fermi Research Alliance, LLC under Contract No. DE-AC02-07CH11359 with the U.S. Department of Energy, Office of Science, Office of High Energy Physics. The United States Government retains and the publisher, by accepting the article for publication, acknowledges that the United States Government retains a non-exclusive, paid-up, irrevocable, world-wide license to publish or reproduce the published form of this manuscript, or allow others to do so, for United States Government purposes.

\bibliographystyle{mnras_2author}
\bibliography{apj-jour,astroref}

\section*{Affiliations}
$^{1}$ Max Planck Institute for Extraterrestrial Physics, Giessenbachstrasse, 85748 Garching, Germany\\
$^{2}$ Universit\"ats-Sternwarte, Fakult\"at f\"ur Physik, Ludwig-Maximilians Universit\"at M\"unchen, Scheinerstr. 1, 81679 M\"unchen, Germany\\
$^{3}$ Department of Physics, Stanford University, 382 Via Pueblo Mall, Stanford, CA 94305, USA\\
$^{4}$ Kavli Institute for Particle Astrophysics \& Cosmology, P. O. Box 2450, Stanford University, Stanford, CA 94305, USA\\
$^{5}$ SLAC National Accelerator Laboratory, Menlo Park, CA 94025, USA\\
$^{6}$ Department of Applied Mathematics and Theoretical Physics, University of Cambridge, Cambridge CB3 0WA, UK\\
$^{7}$ Brookhaven National Laboratory, Bldg 510, Upton, NY 11973, USA\\
$^{8}$ D\'epartement de Physique Th\'eorique and Center for Astroparticle Physics, Universit\'e de Gen\'eve, 24 quai Ernest Ansermet, CH-1211Geneva, Switzerland\\
$^{9}$ Center for Cosmology and Astro-Particle Physics, The Ohio State University, Columbus, OH 43210, USA\\
$^{10}$ Fermi National Accelerator Laboratory, P. O. Box 500, Batavia, IL 60510, USA\\
$^{11}$ Kavli Institute for Cosmological Physics, University of Chicago, Chicago, IL 60637, USA\\
$^{12}$ Argonne National Laboratory, 9700 South Cass Avenue, Lemont, IL 60439, USA\\
$^{13}$ Department of Physics, University of Arizona, Tucson, AZ 85721, USA\\
$^{14}$ Faculty of Physics, Ludwig-Maximilians-Universit\"at, Scheinerstr. 1, 81679 Munich, Germany\\
$^{15}$ Department of Physics and Astronomy, University of Pennsylvania, Philadelphia, PA 19104, USA\\
$^{16}$ Department of Physics, Carnegie Mellon University, Pittsburgh, Pennsylvania 15312, USA\\
$^{17}$ Santa Cruz Institute for Particle Physics, Santa Cruz, CA 95064, USA\\
$^{18}$ Center for Astrophysical Surveys, National Center for Supercomputing Applications, 1205 West Clark St., Urbana, IL 61801, USA\\
$^{19}$ Department of Astronomy, University of Illinois at Urbana-Champaign, 1002 W. Green Street, Urbana, IL 61801, USA\\
$^{20}$ Department of Physics, University of Oxford, Denys Wilkinson Building, Keble Road, Oxford OX1 3RH, UK\\
$^{21}$ Jodrell Bank Center for Astrophysics, School of Physics and Astronomy, University of Manchester, Oxford Road, Manchester, M13 9PL, UK\\
$^{22}$ Centro de Investigaciones Energ\'eticas, Medioambientales y Tecnol\'ogicas (CIEMAT), Madrid, Spain\\
$^{23}$ Department of Physics, Duke University Durham, NC 27708, USA\\
$^{24}$ Institute for Astronomy, University of Edinburgh, Edinburgh EH9 3HJ, UK\\
$^{25}$ Departamento de F\'isica Matem\'atica, Instituto de F\'isica, Universidade de S\~ao Paulo, CP 66318, S\~ao Paulo, SP, 05314-970, Brazil\\
$^{26}$ Laborat\'orio Interinstitucional de e-Astronomia - LIneA, Rua Gal. Jos\'e Cristino 77, Rio de Janeiro, RJ - 20921-400, Brazil\\
$^{27}$ CNRS, UMR 7095, Institut d'Astrophysique de Paris, F-75014, Paris, France\\
$^{28}$ Sorbonne Universit\'es, UPMC Univ Paris 06, UMR 7095, Institut d'Astrophysique de Paris, F-75014, Paris, France\\
$^{29}$ Department of Physics and Astronomy, Pevensey Building, University of Sussex, Brighton, BN1 9QH, UK\\
$^{30}$ Department of Physics \& Astronomy, University College London, Gower Street, London, WC1E 6BT, UK\\
$^{31}$ Instituto de Astrofisica de Canarias, E-38205 La Laguna, Tenerife, Spain\\
$^{32}$ Universidad de La Laguna, Dpto. Astrofísica, E-38206 La Laguna, Tenerife, Spain\\
$^{33}$ Institut de F\'{\i}sica d'Altes Energies (IFAE), The Barcelona Institute of Science and Technology, Campus UAB, 08193 Bellaterra (Barcelona) Spain\\
$^{34}$ Astronomy Unit, Department of Physics, University of Trieste, via Tiepolo 11, I-34131 Trieste, Italy\\
$^{35}$ INAF-Osservatorio Astronomico di Trieste, via G. B. Tiepolo 11, I-34143 Trieste, Italy\\
$^{36}$ Institute for Fundamental Physics of the Universe, Via Beirut 2, 34014 Trieste, Italy\\
$^{37}$ Observat\'orio Nacional, Rua Gal. Jos\'e Cristino 77, Rio de Janeiro, RJ - 20921-400, Brazil\\
$^{38}$ Department of Physics, University of Michigan, Ann Arbor, MI 48109, USA\\
$^{39}$ Department of Physics, IIT Hyderabad, Kandi, Telangana 502285, India\\
$^{40}$ Institute of Theoretical Astrophysics, University of Oslo. P.O. Box 1029 Blindern, NO-0315 Oslo, Norway\\
$^{41}$ Instituto de Fisica Teorica UAM/CSIC, Universidad Autonoma de Madrid, 28049 Madrid, Spain\\
$^{42}$ Institut d'Estudis Espacials de Catalunya (IEEC), 08034 Barcelona, Spain\\
$^{43}$ Institute of Space Sciences (ICE, CSIC),  Campus UAB, Carrer de Can Magrans, s/n,  08193 Barcelona, Spain\\
$^{44}$ Department of Astronomy, University of Michigan, Ann Arbor, MI 48109, USA\\
$^{45}$ School of Mathematics and Physics, University of Queensland,  Brisbane, QLD 4072, Australia\\
$^{46}$ Department of Physics, The Ohio State University, Columbus, OH 43210, USA\\
$^{47}$ Australian Astronomical Optics, Macquarie University, North Ryde, NSW 2113, Australia\\
$^{48}$ Lowell Observatory, 1400 Mars Hill Rd, Flagstaff, AZ 86001, USA\\
$^{49}$ Department of Astrophysical Sciences, Princeton University, Peyton Hall, Princeton, NJ 08544, USA\\
$^{50}$ Instituci\'o Catalana de Recerca i Estudis Avan\c{c}ats, E-08010 Barcelona, Spain\\
$^{51}$ Physics Department, 2320 Chamberlin Hall, University of Wisconsin-Madison, 1150 University Avenue Madison, WI  53706-1390\\
$^{52}$ Institute of Astronomy, University of Cambridge, Madingley Road, Cambridge CB3 0HA, UK\\
$^{53}$ School of Physics and Astronomy, University of Southampton,  Southampton, SO17 1BJ, UK\\
$^{54}$ Computer Science and Mathematics Division, Oak Ridge National Laboratory, Oak Ridge, TN 37831\\
$^{55}$ Institute of Cosmology and Gravitation, University of Portsmouth, Portsmouth, PO1 3FX, UK\\

\appendix

\section{Data Selection}

 The wide-field galaxy sample used in this study for the statistical modeling (\autoref{sec:statmodel}) is obtained from the DES Y3 GOLD galaxy catalog  \citep{Y3gold} using the criteria listed in \autoref{tab:y3mof_cuts}. The full list of galaxy features used in this study are listed in \autoref{tab:feature_list} along with their relation to the DES Y3 data products produced by \citet{Y3gold} and \citet*{Y3deep}, corresponding to the wide-field and deep-field features respectively. The mean photometric and morphological parameters of \redmapper\ BCGs are listed in \autoref{tab:bcgs}. These are obtained by matching the galaxy properties of the Y3 GOLD catalog with the catalog of \redmapper\ central galaxies based on the \texttt{COADD\_OBJECT\_ID}.

\begin{table*}
    \renewcommand{\arraystretch}{1.3}
    \footnotesize
    \begin{tabular}{lll}
    Feature & catalog parameter & description\\
    \hline
    \hline
    Deep-field features\\
    \hline
    $m$  & \texttt{bdf\_mag\_dered\_3} & i-band MOF magnitude with photometric  correction\\
    \hline
    $\vect{c}$  & \texttt{bdf\_mag\_dered\_2} - \texttt{bdf\_mag\_dered\_1} &  g - r MOF color with photometric  correction\\
    & \texttt{bdf\_mag\_dered\_3} - \texttt{bdf\_mag\_dered\_2} & r - i MOF color with photometric  correction\\
    & \texttt{bdf\_mag\_dered\_4} - \texttt{bdf\_mag\_dered\_3} & i - z MOF color with photometric  correction\\     
    
    \hline
    $\vect{s}$ & sqrt(\texttt{bdf\_g\_0}$^2$ +  \texttt{bdf\_g\_1}$^2$)& absolute MOF ellipticity $\left| \vect{e} \right|$\\
    & \texttt{FRACDEV}& bulge / disk flux fraction at fixed component size\\
    & log$_{10}$(1 + \texttt{bdf\_T})&  MOF size squared in  arcsec$^2$ $T=<x^2>+<y^2>$ \\    
    \hline
    $z_g$ & \texttt{z\_mc} & $ugrizJHK$-band based photo-z estimate from EAzY\\
    \hline
    \hline
    Wide-field features\\
    \hline
    $R$ & log$_{10} \sqrt{(\mathrm{\texttt{RA} - ra_{ref}})^2 + (\mathrm{\texttt{DEC} - dec_{ref}})^2}$ & log$_{10}$ projected separation in arcmin from reference point\\
    \hline
    $m$ & \texttt{MOF\_CM\_MAG\_CORRECTED\_I} & i-band MOF magnitude with photometric  correction\\
    \hline
    $\vect{c}$ & \texttt{MOF\_CM\_MAG\_CORRECTED\_G} - \texttt{MOF\_CM\_MAG\_CORRECTED\_R} & g - r MOF color with photometric  correction\\
     & \texttt{MOF\_CM\_MAG\_CORRECTED\_R} - \texttt{MOF\_CM\_MAG\_CORRECTED\_I} & r - i  MOF color with photometric  correction\\
     & \texttt{MOF\_CM\_MAG\_CORRECTED\_I} - \texttt{MOF\_CM\_MAG\_CORRECTED\_Z} & i - z MOF color with photometric  correction\\
    \hline
    \end{tabular}
    \caption[Feature definitions from the column of the relevant photometric catalogs.]{Features and their definitions from the column of the relevant photometric catalogs. Deep  field features: DES Y3 deep and supernova fields \citep*{Y3deep} for further explanation see \autoref{sec:des_deep}. Wide-field features: DES Y3 GOLD \citep{Y3gold}, for further explanation see \autoref{sec:des_wide}.
    }    
	\label{tab:feature_list}
\end{table*}

\begin{table*}
\centering
    \renewcommand{\arraystretch}{1.3}
    \small
    \begin{tabular}{lll}
    Y3A2 GOLD column & value & description\\
    \hline
    \hline
    \texttt{FLAGS\_FOOTPRINT} & 1 & restricts catalog to fiducial survey footprint\\
    \texttt{FLAGS\_FOREGROUND} & 0 & excludes regions masked due to foreground objects\\
    \texttt{bitand(FLAGS\_GOLD, 122)} & 0 & photometric processing failure exclusion based on SOF\\
    \texttt{EXTENDED\_CLASS\_SOF} & 3 & high purity galaxy sample based on SOF model\\
    \hline    
    \end{tabular}
    \caption[Y3A2 GOLD catalog query cuts]{Y3A2 GOLD catalog query cuts used in obtaining the survey data from the DES Data Management System (DESDM, \citealp{DESDM}).} 
	\label{tab:y3mof_cuts}
\end{table*}

\begin{table*}
    \renewcommand{\arraystretch}{1.3}
    \centering
    \begin{tabular}{l l l l l l l l}
    $z\in$ & $\lambda\in$ & $\langle i\rangle$ & $\langle g-r\rangle $ & $\langle r-i\rangle $ & $\langle i-z\rangle $ & $\langle T_{BCG} \rangle$ [arcsec$^{2}$] & $\langle |g|\rangle $\\
    \hline
    $[0.3;\,0.35)$ & $[30;\,45)$ & 17.76 & 1.36 & 0.54 & 0.32 & 28.90 & 0.14\\ 
    $[0.3;\,0.35)$ & $[45;\,60)$ & 17.62 & 1.38 & 0.54 & 0.31 & 33.20 & 0.14\\
    $[0.45;\,0.5)$ & $[30;\,45)$ & 18.58 & 1.85 & 0.70 & 0.37 & 21.92 & 0.15\\
    $[0.45;\,0.5)$ & $[45;\,60)$ & 18.50 & 1.85 & 0.71 & 0.37 & 28.43 & 0.14\\
    $[0.6;\,0.65)$ & $[30;\,45)$ & 19.36 & 1.83 & 1.01 & 0.44 & 16.90 & 0.17\\
    $[0.6;\,0.65)$ & $[35;\,60)$ & 19.18 & 1.83 & 1.02 & 0.45 & 22.44 & 0.16\\
    \hline
    \end{tabular}
    \caption{Properties of the mean bright central galaxy (BCG) across the different cluster richness and redshift bins. For each BCG the bulge (de Vaucouleurs) fraction is set to unity. The $T_{BCG}$ parameter is the effective area of the galaxy corresponding to the SOF size squared in  arcsec$^2$ $T=<x^2>+<y^2>$.}
	\label{tab:bcgs}
\end{table*}

\label{lastpage}
\end{document}